\documentclass[conference,compsoc]{IEEEtran}

\usepackage[T1]{fontenc}
\usepackage[nocompress]{cite}
\usepackage{amsmath,amssymb}
\usepackage{booktabs}
\usepackage{graphicx}
\usepackage{tabularx}
\usepackage{tikz}
\usepackage{xspace}
\usepackage{placeins}
\usepackage{xcolor}
\usepackage{colortbl}
\usepackage{xurl}
\usepackage{hyperref}
\usepackage{enumitem}
\usepackage{graphicx}
\usepackage[caption=false,font=footnotesize,labelfont=sf,textfont=sf]{subfig}

\usepackage{booktabs}  
\usepackage{tabularx}  
\usepackage{xcolor}    
\usepackage{colortbl}  
\usepackage{amsmath}   

\definecolor{DAAccent}{HTML}{243B5A}
\definecolor{DAPale}{HTML}{EAF0F6}
\definecolor{DASoft}{HTML}{F5F7FA}
\definecolor{DAWarn}{HTML}{9A4638}
\hypersetup{
  colorlinks=true,
  linkcolor=DAAccent,
  citecolor=DAAccent,
  urlcolor=DAAccent,
  pdftitle={Are Web Agents Still Detectable? Stress-Testing Bot Defenses with Coordinated Evasion},
  pdfauthor={Parsa Nilforushan, Gemmin Sugiura, Kourosh Hashemi, Peiman Zhiani, Mohammad A Tayebi}
}

\newcommand{\FPAgent}{FP-Agent\xspace}
\newcommand{\GhostCursor}{Ghost Cursor\xspace}
\newcommand{\DarkAgent}{\textsc{DarkAgent}\xspace}
\newcommand{\Browsium}{\textsc{Browsium}\xspace}
\newcommand{\Behavium}{\textsc{Behavium}\xspace}
\newcommand{\MouseGen}{\textsc{MouseGen}\xspace}
\newcommand{\MimicTrace}{\textsc{MimicTrace}\xspace}
\newcommand{\KeyboardGen}{\textsc{KeyboardGen}\xspace}
\newcommand{\researchq}[1]{\textbf{RQ#1}}
\newcommand{\vect}[1]{\mathbf{#1}}
\newcommand{\Normal}{\mathcal{N}}

\title{
    Are Web Agents Still Detectable? \\ Stress-Testing Bot Defenses with Coordinated Evasion
}
\author{
\IEEEauthorblockN{Parsa Nilforushan\textsuperscript{*}, Gemmin Sugiura\textsuperscript{*}, Kourosh Hashemi,\\
Peiman Zhiani, Mohammad A. Tayebi}
\IEEEauthorblockA{
 Computing Science \\
Simon Fraser University\\
British Columbia, Canada\\
\{parsa\_nilforushan, gemmin\_sugiura, kourosh\_hashemi, pza42, tayebi\}@sfu.ca}
}

\begin{document}
\bstctlcite{IEEEtran:BSTcontrol}

\maketitle
\begingroup
\renewcommand{\thefootnote}{\fnsymbol{footnote}}
\footnotetext[1]{Parsa Nilforushan and Gemmin Sugiura contributed equally to this work.}
\endgroup

\begin{abstract}
AI web agents can autonomously navigate websites and complete various tasks. As their capabilities and adoption expand, evaluating bot defenses requires adversarial stress tests. Although prior work demonstrates that existing agents can remain detectable, it remains unclear how effective bot defenses are against adversaries that coordinate evasion across browser fingerprints, automation controls, and interaction behavior. To investigate this threat, we introduce \DarkAgent, an autonomous web agent that combines a forked Chromium browser that conceals automation artifacts and generative models that produce human-like mouse and keyboard traces. To address the limitations of existing keyboard mimickers, our keyboard model jointly generates realistic key events and their timings while reliably reproducing the target text. Released agent detectors accept our generated behavior as human, and trained classifiers find it substantially harder to distinguish from human traces than existing mimickers. \DarkAgent achieves an 89\% page-access bypass rate across a third-party stealth benchmark, 
exceeding the strongest published baseline by 12 percentage points. To further enhance the measurement of agents, we introduce a benchmark that measures browser fingerprints, interaction behavior, and network activity to assess each channel's contribution to detecting adversarial agents such as \DarkAgent. On this benchmark, browser and behavioral modifications are complementary: either alone remains detectable, and only their combination evades all tested checks. Together, our findings expose the gap between passing existing bot defenses and reproducing human behavior, and point to timing between actions as a remaining signal for defenders.
\end{abstract}
\section{Introduction}
\label{sec:introduction}

Web agents powered by multimodal language models can interpret interfaces,
plan actions, and complete online tasks from natural-language
instructions~\cite{he2024webvoyager}. Their ability to adapt to unfamiliar
interfaces makes useful automation more accessible, but lowers the
barrier to automating abuse, including mass account creation, spam, and
unauthorized scraping~\cite{agentshield}. Website operators thus need
to identify automated sessions and enforce access policies even when agents
use the same browsers and interfaces as legitimate users.

Anti-bot defenses examine network characteristics, browser fingerprints,
automation artifacts, and interaction
behavior~\cite{cloudflareBotEngines,agentshield}. These signals arise from
different parts of the client stack. A browser can expose automation
properties independently of the mouse movements or keystrokes it
delivers~\cite{webdriver}; conversely, suppressing browser artifacts does
not make mechanical input resemble human behavior. Anti-bot robustness
therefore needs to be evaluated against coordinated changes to the browser
and its interactions. Recent studies establish that existing web agents exhibit detectable
behavioral differences~\cite{wang2026fpagent,minimalfeatures,agentshield}.
AgentShield evaluates agents prompted to imitate humans, while identifying
fully custom agents as a direction for future work~\cite{agentshield}.

Meaningful evaluation of defenses requires sufficiently capable
adversaries. Otherwise, strong detection performance may reflect limitations
of the attacks being evaluated. Public tools already support browser
concealment and behavioral
mimicry~\cite{ghostcursor,humantyping,cloakbrowser}, making coordinated
evasion a practical threat. Choudhary et
al.~\cite{minimalfeatures} traced detection to missing raw pointer events
in Playwright sessions with generated mouse trajectories or replays of
human movements. This leaves open whether defenses can distinguish
generated behavior from human input when browser-control and input-delivery
artifacts are absent and browser fingerprints agree across channels. For keyboard behavior, prior work includes timing
generators, state-based imitation, and heuristic editing
simulations~\cite{becaptchatype,keygan,pashchenko2020imitation,humantyping}.
However, the joint problem of modeling text-conditioned physical key events,
including mistakes, corrections, and overlapping key holds, together with
their timing remains comparatively underexplored. Together, these limitations motivate evaluating defenses against a more capable adversary that coordinates browser concealment with learned behavioral generation.

Thus, we introduce \DarkAgent, a modular autonomous web agent for investigating
this threat. \DarkAgent combines \Browsium, our stealthy browser forked from
Chromium to conceal automation artifacts, with \Behavium, our keyboard and
mouse generation layer.
\Browsium separates agent control from renderer DevTools attachment through
a private communication channel, observes pages through isolated worlds,
and delivers input through the browser's native mechanisms. \Behavium includes \KeyboardGen, which models the joint distribution of physical key events and their timing conditioned on target text, using two autoregressive Transformers coupled through a shared latent representation of typing style. 
The high-level planner selects actions, \Behavium generates their input sequences, and
\Browsium executes them. The browser and behavioral components can
be varied independently, enabling controlled comparisons of their
individual and combined effects.

Four research questions guide our investigation:

\label{sec:rqs}
\begin{itemize}[leftmargin=*]
  \item \researchq{1}: \textbf{Detection effectiveness.} How effective are
  existing anti-bot defenses against agents using stealthy browsers and
  behavioral mimicry?
  \item \researchq{2}: \textbf{Component contributions.} How do browser,
  mouse, and keyboard modifications contribute to evasion, individually
  and together?
  \item \researchq{3}: \textbf{Behavioral fidelity.} How distinguishable
  are generated mouse and keyboard interactions from human behavior,
  compared with existing mimicry methods?
  \item \researchq{4}: \textbf{Detection generalization.} How well does detection based on behavioral and execution-environment evidence generalize to unseen devices and agent families?
\end{itemize}

Across the 80-task Stealth Bench~V1
suite~\cite{stealthbenchv1,stealthbenchv1repository}, \DarkAgent achieves an
89\% bypass rate, while controlled ablations reveal complementary
browser and behavioral detection signals. However, detector acceptance alone does not establish behavioral resemblance or generalization to unseen agents. To investigate these limitations, we introduce \MimicTrace, a benchmark that captures browser, network, automation-control, and behavioral signals alongside server-verified task outcomes.

This paper makes the following contributions:
\begin{itemize}[leftmargin=*]
  \item \textbf{An architecture for coordinated evasion.}
  We develop \DarkAgent and \Browsium to combine browser-level changes
  with native execution of generated input, enabling controlled variation
  of the browser, mouse, and keyboard components.

  \item \textbf{Generative input with explicit functional requirements.}
  We introduce \KeyboardGen for state-aware event and timing generation,
  and \MouseGen for conditional trajectory generation, separating
  functional correctness from resemblance to human behavior.

  \item \textbf{Detection limits and component effects.}
  Through component ablations and comparisons with human traces and public
  mimickers, we identify complementary detection signals, failures to
  generalize across input generators, and residual differences in accepted
  sessions.
  \item \textbf{A benchmark for multi-channel detection and generalization.}
  We introduce \MimicTrace, which links browser, network, control, and
  behavioral telemetry with server-verified task outcomes. It supports
  joint analysis of residual detection signals and evaluates whether
  detection generalizes to unseen devices and agent families.
\end{itemize}

\section{Background and Related Work}
\label{sec:related-work}

\subsection{Browser Automation and Detection Surfaces}
\label{sec:automation-background}

An autonomous web agent repeatedly observes a page, selects an action,
and executes it through a browser. Its planning policy determines the
action and target; its execution stack determines how page state is read
and how input reaches the page. Automation frameworks use mechanisms
such as the Chrome DevTools Protocol (CDP), injected scripts, and browser
input facilities~\cite{playwrightCDPSession,playwrightPageAPI,playwrightActions}. These mechanisms can expose different artifacts even
when they produce the same visible page
state~\cite{agentshield,patchright}.
Appendix~\ref{sec:background} describes the relevant browser boundaries.

We distinguish four channels of detection evidence.
\emph{Automation-control} artifacts expose instrumentation or injected
helpers, while \emph{browser} fingerprints describe properties such as
the reported platform, supported APIs, and rendering behavior.
\emph{Network} evidence includes TLS and HTTP characteristics;
\emph{behavioral} evidence includes mouse motion, scrolling, keystrokes,
and interaction timing~\cite{li2021goodbot,fayolle2026llmbot,wang2026fpagent,agentshield}.
Defenders can inspect each channel and test consistency between them,
such as agreement between browser-reported properties and request
headers. This motivates evaluating behavioral generation together with
the browser mechanisms that execute it.

\subsection{Threat Model}
\label{sec:threat-model}
We consider an attacker-controlled autonomous web agent interacting
with a website protected by bot defenses. The attacker's objective is
to complete the requested task correctly while avoiding restrictions
on automated activity.

\noindent\textbf{Attacker.}
The attacker controls the client stack, including the browser,
automation mechanisms, and behavioral generation. It can modify browser
source code, change how pages are observed and controlled, and deliver
behavioral model-generated input through native
browser mechanisms. It may inspect content and scripts delivered to
the browser, observe website responses, and adapt the client
accordingly. The system grants no privileged access to private
server-side state or detection logic, and no direct control over
enforcement decisions.

\noindent\textbf{Defender.}
The defender is the website operator or a bot-protection service acting
on its behalf. It seeks to enforce its policy on automated activity
while permitting legitimate human use. It can instrument pages, probe
browser and automation artifacts, record interactions, and combine
these observations with network, request, and session-level signals.
It may change page content, issue challenges, or block access in response
to suspicious activity. Privileged inspection of
the agent's host is outside this model.

\noindent\textbf{Knowledge and measurement scope.}
The defender's capabilities are independent of the signals recorded by
a particular benchmark. Access to public detector implementations or
self-hosted audits does not imply access to a deployed service's private
systems. We distinguish live-site blocking, controlled detector
decisions, behavioral fidelity, and verified task completion: acceptance
by a detector alone establishes neither task correctness nor
resemblance to human behavior.

\subsection{Stealthy Browsers}
\label{sec:related-stealth-browsers}
Stealthy browsers are modified versions of browsers such as Chromium or
Firefox, designed to reduce browser and automation-control evidence used by
bot detectors. Automation patchers instead modify the frameworks that
control browsers. Patchright~\cite{patchright}
modifies Playwright to reduce CDP-related leaks, while
Camoufox~\cite{camoufox} modifies Firefox's fingerprinting surfaces.
Browser Use Cloud~\cite{browserusecloud} and
CloakBrowser~\cite{cloakbrowser} use modified Chromium builds to alter
browser fingerprints.

\subsection{Keyboard Mimicry}
\label{sec:keyboard-mimicry}

\noindent\textbf{Keyboard behavior.}
Concealing runtime artifacts still leaves automated input behavior observable~\cite{agentshield}. Prior work generates keyboard behavior for biometric forgery~\cite{syntheticforgery}, bot detection, and digital phenotyping. BeCAPTCHA-Type~\cite{becaptchatype} reconstructs key-press and key-release timestamps for supplied key sequences, while KeyGAN~\cite{keygan} generates timing sequences without retaining key identities. These tasks differ from learning how to type requested text: the target specifies the intended result, but leaves open the event sequence used to produce it, including any mistakes and corrections. \KeyboardGen learns both the physical key-down/key-up sequence and its timing through two autoregressive decoders conditioned on a shared latent variable. It models mistakes, corrections, and overlapping key holds (rollover) while enforcing legal key transitions; successful completion of the requested text is evaluated empirically (Section~\ref{sec:keyboard-generation}).

\subsection{Web-Agent Defenses and Benchmarks}
\label{sec:related-detection}
Web-agent defenses include measures that disrupt content extraction
and mechanisms that classify automated sessions.
WebCloak~\cite{li2026webcloak} impedes LLM-driven scraping through
dynamic structural obfuscation and semantic labyrinths. Commercial systems such as
Cloudflare~\cite{cloudflareBotEngines} combine request, browser, and
session signals, with detection engines varying by deployment.
Access restrictions and deployment-specific configurations complicate
reproducible stress-testing of these systems.

\noindent\textbf{Agent detection.}
Multi-layer fingerprinting~\cite{fayolle2026llmbot,kang2026whose} examines network,
HTTP, and browser artifacts.
\FPAgent~\cite{wang2026fpagent} combines browser fingerprints with
50 behavioral features. Its published features omit cursor speed,
acceleration, and dedicated rollover statistics.
AgentShield~\cite{agentshield} uses dynamic forms to expose movement
volume, keypress durations, and field-reaction latencies, and evaluates
agents prompted to imitate human behavior.
Minimal-feature behavioral detection~\cite{minimalfeatures} tests
generated cursor trajectories and human replay within
Playwright-mediated execution.
\DarkAgent extends this evaluation to client stacks that
coordinate browser isolation with learned mouse and keyboard
generation.

\noindent\textbf{Benchmarks and measurement.}
Stealth Bench~V1~\cite{stealthbenchv1,stealthbenchv1repository}
measures blocking on live sites, but its verdict neither attributes
blocking to a detection channel nor independently verifies task
correctness. Human interaction datasets support detailed mouse or
keyboard analysis, but do not jointly capture browser, network, and
interaction evidence across complete browsing
sessions~\cite{dhakal2018typing,kilic2021bogazici,leiva2020attentive}.
\MimicTrace addresses this measurement gap: its collector records all
four evidence channels within each session, alongside server-verified
task completion.

\section{Methodology}
\label{sec:methodology}

\begin{figure*}[t]
\centering
\includegraphics[width=0.90\textwidth]{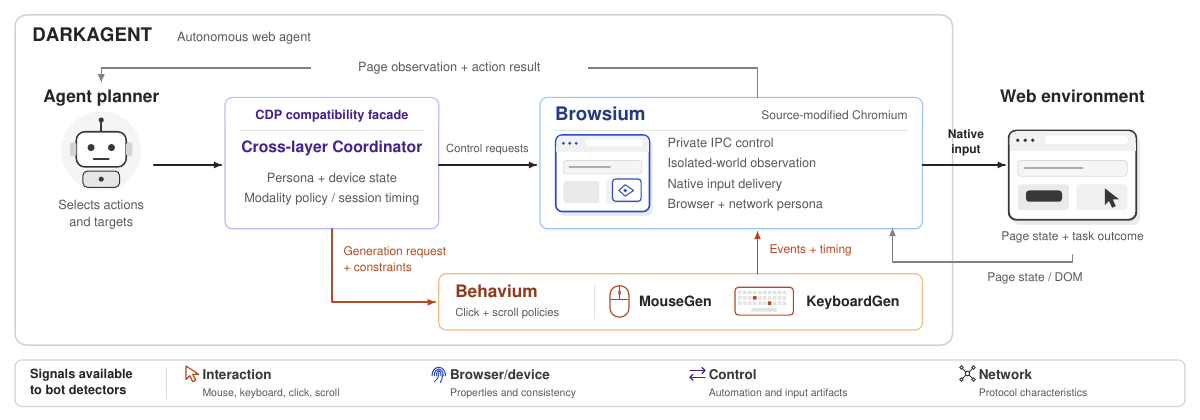}
\caption{\textbf{\DarkAgent architecture.} The planning agent selects actions
from structured page observations. \DarkAgent routes requests through a
compatibility interface over \Browsium's private IPC channel, pairs them with
mouse and keyboard behavior from \Behavium, and executes them via native input
delivery before returning updated page observations.}
\label{fig:architecture}
\end{figure*}

\DarkAgent coordinates evasion across the automation-control, browser, network,
and behavioral channels. A high-level planning
agent selects actions and targets from page observations, which contain
structured DOM information and, when needed, a screenshot
(Figure~\ref{fig:architecture}). \Browsium executes page
observation and input without a renderer DevTools attachment, while \Behavium
generates mouse trajectories, physical key-event sequences, and click and
scroll timings. Each component can be toggled independently to isolate its
contribution to evasion.

\noindent\textbf{Motivating example.}
Consider an agent submitting an inquiry form whose submit button sits inside a
closed shadow root on a protected website. Conventional automation introduces
automation artifacts via CDP or main-world scripts when inspecting that button
and emits mechanical detection signals when typing or clicking
(Appendix~\ref{sec:background}). \DarkAgent separates planning from execution:
the planning agent selects the action; \Browsium resolves the closed-shadow
button in its hidden isolated world; \Behavium generates a timed trajectory or
physical key-event sequence; and \Browsium delivers the events through native
input delivery before returning an updated observation.

\subsection{Stealthy Browser Architecture (Browsium)}
\label{sec:stealth-browser architecture}

\begin{figure*}[t]
    \centering

    \subfloat[CDP-based browser automation.\label{fig:cdp-automation}]{%
        \includegraphics[width=0.90\textwidth]
            {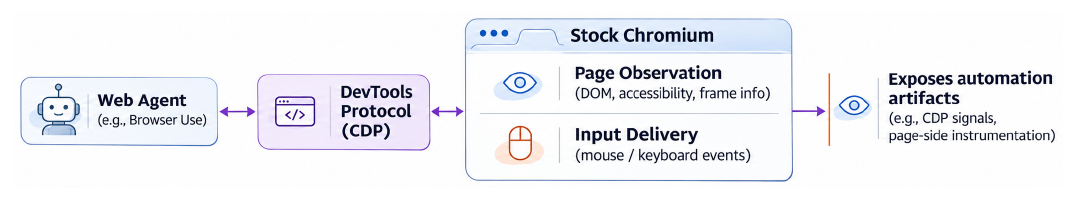}%
    }

    \par\medskip

    \subfloat[\Browsium.\label{fig:browsium-control}]{%
        \includegraphics[width=0.90\textwidth]
            {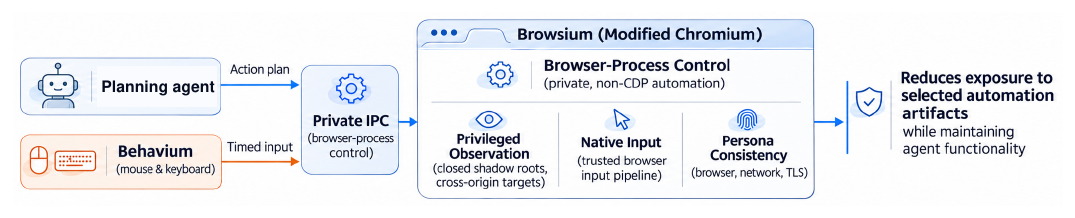}%
    }

    \caption{\textbf{Browser-control architectures.}
    (a) CDP-based automation uses the DevTools protocol for page
    observation and input delivery in stock Chromium.
    (b) \Browsium uses private browser-process control to support
    privileged observation, native input delivery, and browser/network
    persona consistency, with \Behavium supplying timed mouse and
    keyboard interactions.}
    \label{fig:browsium-architecture}
\end{figure*}

Web agents need privileged access to inspect page state and trigger input, yet
conventional frameworks attach DevTools to the renderer or inject main-world
scripts, introducing observable automation artifacts
(Appendix~\ref{sec:background}). Existing stealth tools patch these artifacts
disjointly, leaving contradictions across execution worlds and detection
channels. We instead draw inspiration from Electron's context
isolation~\cite{electronContextIsolation}, which separates privileged scripts
from untrusted page code. Electron's \texttt{contextBridge} can explicitly expose APIs
in the main world, and standard Blink isolated worlds can neither
inspect closed shadow roots nor dispatch trusted input. \Browsium's
architectural contribution is \emph{asymmetric, one-way isolation}
(Figure~\ref{fig:browsium-architecture}): page observations flow \emph{upward}
from a hidden isolated world with zero main-world writes, while control and
native input delivery flow \emph{top-down} through the browser process. Four
design choices realize this architecture across the browser's boundaries:

\noindent\textbf{Out-of-renderer browser control.}
CDP can use a debugging port or a pipe, but pipe transport alone does not
avoid renderer inspector attachment or \texttt{Runtime.enable} serialization
side effects that page scripts can provoke (Appendix~\ref{sec:bg-control}). Rather than suppressing individual CDP
domains, \Browsium decouples control from the renderer by terminating an
operating-system private IPC channel at a command dispatcher in the browser
process. A compatibility interface maps requests from the planning agent onto
this channel (Section~\ref{sec:coordinated-evasion}), eliminating localhost debugging
sockets, \texttt{navigator.webdriver}, and renderer inspector attachment,
though it does not alter browser fingerprints or interaction timing.

\noindent\textbf{Asymmetric isolated-world observation.}
Agents must resolve targets inside closed shadow roots and cross-origin
out-of-process iframes (OOPIFs), yet main-world helper scripts alter prototypes
and globally opening closed shadow roots fails main-world defender self-tests
(Appendix~\ref{sec:bg-isolation}). \Browsium instead enforces world-gated
visibility inside a dedicated hidden isolated world. Blink exposes closed
shadow roots only to that hidden world while preserving
\texttt{Element.shadowRoot == null} in the main world, and browser-process Mojo
queries compose cross-frame bounding boxes without running main-world
JavaScript. This yields full target visibility with zero main-world footprint,
though observation alone cannot emit trusted user input.

\noindent\textbf{Top-down native input delivery.}
Events synthesized inside any renderer world carry \texttt{Event.isTrusted =
false} and bypass compositor hit-testing, while headless windows often lack
active focus (Appendix~\ref{sec:bg-isolation}). Instead of dispatching events
from JavaScript, \Browsium routes actions through the browser process's native
input router and maintains active focus in the focus controller. Mouse,
keyboard, and wheel sequences traverse normal compositor transforms and OOPIF
hit-testing to arrive with \texttt{Event.isTrusted = true} and
\texttt{document.hasFocus() == true} without a DevTools session, while
\Behavium supplies the trajectories and timings carried on this path.

\noindent\textbf{Single-source cross-channel persona.}
Defenses cross-check JavaScript device APIs against HTTP headers, TLS
handshakes, and WebRTC routes, catching surface-level getter overrides
(Appendix~\ref{sec:bg-channels}). Rather than patching APIs in the main world,
\Browsium derives one session persona from a startup seed and applies it across
Chromium's networking and rendering layers. Wire-level TLS \texttt{ClientHello}
parameters, ordered HTTP and \texttt{Sec-CH-UA*} headers, and suppressed
non-proxied WebRTC UDP probes match the same Chrome version and OS reported by
Blink's navigator, display, font, graphics, and media surfaces. This eliminates
cross-channel contradictions within the configured Chromium profile, while IP
reputation and server-side signals remain outside client scope.

Together, these four design choices enforce one-way page observation and
top-down native execution across the automation-control, browser, and network
channels, complementing \Behavium in the behavioral channel
(Section~\ref{sec:behavorial-generation}).

\subsection{Behavioral Generation (Behavium)}
\label{sec:behavorial-generation}

We learn a distribution over keyboard and mouse traces, focusing on keyboard generation to address the asymmetry in prior work. \Behavium must satisfy two complementary criteria: \textit{behavioral fidelity}, or resemblance to human behavior, and \textit{functional correctness}, or compliance with physical and task constraints. For example, a keyboard sequence must produce the target text exactly, obey key-down and key-up constraints, and allow natural mistakes and corrections. Neither criterion implies the other. The datasets and data preparation are specified in Section~\ref{sec:behavioral-setup}.

\subsubsection{Keyboard Generation (\KeyboardGen)}
\label{sec:keyboard-generation}
\mbox{}\par\nobreak
\vspace{0.5\baselineskip}

\noindent\textbf{Formulation.}
Let $\vect x=(x_1,\ldots,x_N)$ be a target text (e.g., "hello world.", "CAT"). The event sequence is $\vect e=(e_1,\ldots,e_T)$, with each $e_i\in\mathcal E$ naming a key and a down or up action, where the first event is placed at time zero. Each event $e_i$ is physically constrained by the preceding events $e_{<i}$. For example, a key cannot be released unless it is currently held, nor can an already-held key be pressed again. However, these constraints do not prohibit multiple keys from being held simultaneously, allowing rollover and modifier combinations such as Shift with a character key. Rollover is particularly important because it occurs frequently in typing, especially among faster typists \cite{dhakal2018typing}. The timing sequence is $\boldsymbol\Delta=(\Delta_2,\ldots,\Delta_T)$, where $\Delta_i$ is the timing gap between $e_{i-1}$ and $e_i$ in milliseconds. We use $i$ for sequence positions and $t$ for the timing gaps. We define the state before an event $e_i$ to be the tuple of the matching prefix length, the divergent suffix length, and the held keys. The convention is $\vect s_i\rightarrow e_i\rightarrow\vect s_{i+1}$, and the initial state $s_1$ has no characters and no held keys.

To model the joint distribution of events and timing gaps, we factorize generation through a shared latent z:
\begin{align}
\label{eq:keyboard-decomposition}
p(\vect e,\boldsymbol\Delta\mid\vect x)
    &= \int p(\vect z) \underbrace{p(\vect e\mid\vect x,\vect z)}_{\text{event}} \underbrace{p(\boldsymbol\Delta\mid\vect e,\vect z)}_{\text{timing gaps}} d\vect z.
\end{align}

\noindent\textbf{Shared latent.}
Generated events and timing gaps should reflect a consistent typing style, including how mistakes, corrections, and pauses occur together. We show that a detector can indeed exploit this inconsistency (Section~\ref{sec:indepedent-behavioral-model-checks}). Thus, to ensure consistency, the latent is \textit{shared} for both generation tasks. During training, a latent $\vect z$ is learned with a Variational AutoEncoder (VAE)~\cite{vae} by encoding the observed events and timing gaps via a bidirectional transformer trained with a weighted KL loss $\lambda_{\mathrm{KL}}\mathcal L_{\mathrm{KL}}$. During generation, we sample $\vect z \sim p(\vect z)$ from the prior, with $p(\vect z)=\Normal(\vect0,\vect I)$. Both decoders are conditioned by a shared latent via adaptive Layer Normalization (AdaLN)~\cite{adaln} within their decoder blocks.

\noindent\textbf{Event generation.}
We model the event-sequence distribution using an autoregressive Transformer encoder-decoder \cite{transformers}. A successful sequence must satisfy: \textit{target-text constraints} and \textit{physical constraints}. Physical constraints require legal key-down and key-up transitions, while permitting overlapping key holds (rollover) and modifier combinations (shift). Target-text constraints require that the event sequence produce the requested text exactly, with all keys released, while still allowing mistakes and corrections along the way.

First, to address the target-text constraint, a bidirectional Target Encoder maps the target text $\vect x$ to contextual embeddings $\vect H_x$, which are computed once and reused throughout generation. Then, at each step $i$, the Event Decoder combines the projected typing state $\vect s_i$ with the previous-event embedding and a position embedding. Conditioned on the shared latent $\vect z$, it applies causal self-attention and cross-attention to $\vect H_x$ to produce logits $\ell_i$ over keyboard events. The generation is successful when the target text is reproduced exactly and all keys are released. However, conditioning on the target text does not guarantee successful generation, which we evaluate empirically in Section~\ref{sec:indepedent-behavioral-model-checks}.

Second, to address physical keyboard constraints, the model both learns and structurally enforces them. Using the current typing state $\vect s_i$, we construct a mask that excludes illegal events while allowing mistakes and corrections. Let $\mathcal E'_i\subseteq\mathcal E$ denote the permitted events at step $i$. Then, the event-sequence distribution is
\begin{equation}
\label{eq:keyboard-event}
p(\vect e\mid\vect x,\vect z) = \prod_{i=1}^{T}\frac{\exp\ell_i(e_i)}{\sum_{e'\in\mathcal E'_i}\exp\ell_i(e')}.
\end{equation}
During training, masked cross-entropy $\mathcal L_{\mathrm e}$ learns the relative probabilities of permitted events but does not directly penalize the logits of illegal events, since they are excluded from the softmax denominator. To provide this learning signal, the auxiliary constraint loss $\mathcal L_{\mathrm c}$ is defined as the average negative log of the total probability mass assigned to permitted events under the unmasked softmax.

\noindent\textbf{Timing generation.}
Once the event sequence is generated, we model the timing-gap-sequence distribution using an autoregressive Transformer Encoder-Decoder. First, for each gap between consecutive events $e_{i-1}$ and $e_i$, the two event embeddings are combined with the typing state $\vect s_i$ and a transition class $c_i$. The transition class $c_i$ is one of \texttt{DOWN--DOWN}, \texttt{DOWN--UP}, \texttt{UP--DOWN}, or \texttt{UP--UP}. A bidirectional encoder maps these features to contextual event-sequence embeddings $\vect H_e$. The autoregressive Timing Decoder uses causal self-attention over previous gaps and cross-attention to $\vect H_e$, conditioned on the shared latent $\vect z$.

Log-normal distributions have been used to model the heavy-tailed distribution of keystroke time intervals~\cite{monaco-2015}. We use the output after cross-attention and transition-class embeddings $c_i$ to parameterize a shared $K$-component log-normal mixture with weights $\pi_{i,k}$, log-space means $\mu_{i,k}$, and positive log-space standard deviations $\sigma_{i,k}$. Since the dataset records event timestamps in milliseconds~\cite{dhakal2018typing}, we model their probabilities using 1-ms quantization cells, which we show empirically improves behavioral fidelity over a continuous likelihood objective (Section~\ref{sec:architectural-justification}). Specifically,
\begin{equation}
\label{eq:keyboard-timing-gaps}
\begin{aligned}
p(\boldsymbol\Delta\mid\vect e,\vect z)
    &=\prod_{i=2}^{T}\sum_{k=1}^{K}\pi_{i,k}\int_{\mathcal Q(\Delta_i)}\frac{1}{t}\mathcal N(\log t;\mu_{i,k},\sigma_{i,k}^2)\,dt.
\end{aligned}
\end{equation}
where $\mathcal Q(\Delta_i)$ is the quantization interval corresponding to the observed gap. We train the Timing Decoder by minimizing the resulting negative log-likelihood $\mathcal L_{\mathrm t}$, computed using differences of log-normal cumulative distribution functions (CDFs).

\noindent\textbf{Training objective.}
We train both decoders using teacher forcing. For a latent $\vect z$, we optimize the objective consisting of the masked event likelihood, auxiliary constraint penalty, quantized log-normal timing likelihood, and KL divergence between the approximate posterior and prior. We have the following objective:
\begin{equation}
\label{eq:keyboard-loss}
\mathcal L=
    \lambda_{\mathrm e}\mathcal L_{\mathrm e}
    +\lambda_{\mathrm c}\mathcal L_{\mathrm c}
    +\lambda_{\mathrm t}\mathcal L_{\mathrm t}
    +\lambda_{\mathrm{KL}}\mathcal L_{\mathrm{KL}}.
\end{equation}

\subsubsection{Mouse Generation (\MouseGen)}
\label{sec:mouse-generation}
\mbox{}\par\nobreak
\vspace{0.5\baselineskip}

\noindent\textbf{Formulation.}
The first objective is to model cursor movements. Let $\mathcal V=[0,W]\times[0,H]$ denote the viewport, $\vect p_0=(x_0,y_0)\in\mathcal V$ the starting cursor position, $\vect p^\star=(x^\star,y^\star)\in\mathcal V$ the target position, and $\vect d=\vect p^\star-\vect p_0$ the net displacement. Given sequence length $L$, let $\vect y=(\vect y_1,\ldots,\vect y_L)$ denote the trajectory with $\vect y_i=(\log\Delta t_i,\Delta x_i,\Delta y_i)$, where $\Delta t_i$ is the timing gap before step $i$ and $\Delta\vect p_i=(\Delta x_i,\Delta y_i)$ is the spatial increment. From $\vect y$, the cursor positions can be constructed via $\vect p_i = \vect p_0 + \sum_j^i \vect \Delta p_j$. Our objective is to generate $\vect y$ such that the final cursor position equals the target ($\vect p^\star = \vect p_L$) and every cursor position remains within the viewport ($\vect p_i \in \mathcal V$). Mirroring the main keyboard Equation~\ref{eq:keyboard-decomposition}, the problem distribution is the following:
\begin{align}
\label{eq:mouse-decomposition}
p(\vect y\mid \vect d, L)
    &= \int p(\vect z) \underbrace{p(\vect y\mid\vect d, L, \vect z)}_{\text{trajectory}} d\vect z,
\end{align}
where $\vect z$ is a latent. We sample the sequence length $L$ from the training displacements.

\noindent\textbf{Trajectory generation.}
Unlike the keyboard model, which generates the event sequence before its timing gaps, the mouse model generates each timing gap followed by its spatial increments. In other words, for a step $i$, we have $\Delta t_i \rightarrow (\Delta x_i, \Delta y_i) \rightarrow \Delta t_{i+1}$, before proceeding to the next $i+1$ step. The model is an autoregressive Transformer conditioned on $\vect d$, $L$, and $\vect z$ via AdaLN.

\noindent\textbf{Functional correctness.}
Prior mouse-generation methods incorporate endpoint information through different mechanisms: DMTG~\cite{liu2024dmtg} initializes trajectories with specified endpoints, BeCAPTCHA-Mouse's knowledge-based generator fits analytic curves through endpoints~\cite{acien2022becaptcha}, and SapiAgent learns to transform start-to-target reference paths into human-like trajectories~\cite{sapiagent}. In contrast, we use a parameter-free, architecture-agnostic spatial increment projection that enforces exact target arrival independently of the underlying generative model.

Specifically, we project generated spatial increments onto the affine constraint set $\mathcal A_{\vect d}=\{(\vect u_1,\ldots,\vect u_L):\sum_{k=1}^{L}\vect u_k=\vect d\}$. The projection has the closed-form solution:
\begin{equation}
\label{eq:mouse-projection}
    \Delta\hat{\vect p}'_k = \Delta\hat{\vect p}_k +\frac{1}{L}(\vect d-\sum_{j=1}^{L}\Delta\hat{\vect p}_j).
\end{equation}
This distributes the endpoint error equally across all increments, guaranteeing exact target arrival while minimizing their total squared adjustment. The projection is differentiable, allowing behavioral-fidelity objectives evaluated on corrected trajectories to be optimized through the correction rather than treating endpoint enforcement as a separate post-processing operation.

For integer-pixel execution, we round cumulative projected positions and
obtain executed increments from consecutive rounded positions, preserving
exact arrival at integer-valued targets. We redistribute canceling unit steps
to avoid zero movements without changing the total displacement. During
training, we use straight-through gradients through rounding. Trajectories
that violate viewport bounds are resampled within a fixed budget rather than
clipped.

\noindent\textbf{Training objective.}
We train \MouseGen as a conditional VAE~\cite{sohn2015}. During training, a bidirectional Transformer encoder maps observed trajectories to an approximate latent posterior. The training objective combines masked categorical cross-entropy over timing gaps and spatial increments with KL regularization. Similar to \KeyboardGen, we model these quantities discretely to match the integer-millisecond and integer-pixel resolution of observed traces.

\subsection{Coordinated Evasion (DarkAgent)}
\label{sec:coordinated-evasion}

A stealthy browser and a behavioral generator are not sufficient on their own during autonomous browsing. \DarkAgent addresses this by separating \emph{what} action to take from \emph{how} that action is executed in the browser. \DarkAgent then coordinates \Behavium and \Browsium to carry out that decision, maintaining coordinated evasion across the automation-control, browser, network, and behavioral channels throughout the session.

\noindent\textbf{Compatibility interface.}
\DarkAgent decouples the planning agent from the browser through a
compatibility interface that provides the same control and observation
functionality as CDP, but reaches \Browsium over the private IPC channel
rather than a DevTools session (Section~\ref{sec:stealth-browser architecture}). The interface
accepts standard CDP-style requests from the planning agent and translates
them into commands for \Browsium's browser-process command dispatcher.
Target and execution-context state expected by the planning agent is
maintained within the interface without attaching a DevTools session to the
renderer. Observation requests are routed to \Browsium's hidden isolated
world and returned in the expected response structures.

\noindent\textbf{Coordinated execution.}
Interaction requests are routed to \Behavium for behavioral generation and
then forwarded over \Browsium's private IPC channel for native input
delivery. For example, when a typing request specifies a target field and
target text, \KeyboardGen generates the physical key-down and key-up event
sequence and its timing gaps, which \Browsium executes through its native
input router. Mouse movement, clicking, and scrolling similarly use
\MouseGen and the click and scroll policies. If generation exhausts its
budget, a path leaves the viewport, or a target disappears, \DarkAgent
returns an action failure; the complete configuration never uses direct DOM
actions or JavaScript-dispatched events as fallbacks. Updated page
observations and action outcomes are then returned to the planning agent to
inform its next step.

\noindent\textbf{Shared session state.}
At session start, \DarkAgent initializes \Browsium's persona and
\Behavium's random state and click profile from the session seed. It
supplies the persona's viewport dimensions to \Behavium and maintains a
shared cursor position across actions, so generated mouse paths start from
the current cursor location and stay within the viewport bounds. Whenever
the page navigates or an action changes page state, \DarkAgent discards
cached page observations and collects a new page observation before the next
action. Browser and behavioral components can also be enabled or disabled
independently at the compatibility interface for the controlled comparisons
in \researchq{2}.

\section{Experimental Setup}
\label{sec:experimental-setup}

For \researchq{1}--\researchq{3}, we describe the agent-level setup
(Section~\ref{sec:agent-level-setup}) and the datasets and tests for
standalone behavioral evaluation (Section~\ref{sec:behavioral-setup}).
Section~\ref{sec:mimictrace} presents the \MimicTrace setup for \researchq{4}.

\subsection{Agent-Level Setup}
\label{sec:agent-level-setup}
All agent runs use the planner and execution interface in
Section~\ref{sec:coordinated-evasion}.

\noindent\textbf{Deployed defenses (\researchq{1}).}
Stealth Bench~V1 uses 80 short, unauthenticated mouse and keyboard tasks across
11 defense families on live websites~\cite{stealthbenchv1,stealthbenchv1repository},
whereas V2 checks whether usable site content becomes accessible within 60
seconds~\cite{stealthbenchv2protocol}. On August~21, 2026, we ran
\DarkAgent across all 80 V1 tasks with \texttt{gpt-4.1} and geo-matched
residential proxies. The V1 judge scores page access only: it counts page-load
failures as blocks and does not check answers or task completion. We define
the bypass rate as the fraction of task runs judged unblocked and report it
with a 95\% Wilson confidence interval alongside all
complete released baseline runs, which the benchmark authors conducted separately on a different date and which we include as reference context.

\noindent\textbf{Component comparisons (\researchq{2}).}
On the self-hosted browsing site, we scheduled 30 sessions for each of the 18
configurations in Table~\ref{tab:matrix} under a fixed planner, prompt, task
(cursor movement, clicking, typing, and scrolling), and step budget. The main
$2\times 2$ comparison crosses stock Chromium and \Browsium with default input
and full \Behavium; Section~\ref{sec:informed-defender} adds headful nodriver
Chrome for a $3\times 2$ comparison. Remaining configurations cover Firefox,
Camoufox~\cite{camoufox}, Patchright~\cite{patchright},
CloakBrowser~\cite{cloakbrowser}, \MouseGen-only and \KeyboardGen-only
ablations, and public mimickers (\GhostCursor~\cite{ghostcursor},
HumanTyping~\cite{humantyping}, and CloakBrowser + Humanize).
Each configuration tests the browser stack as a unit rather than
individual \Browsium patches.

\noindent\textbf{Detection measures.}
We run the released \FPAgent \texttt{behavioral\_fingerprint}
checkpoint~\cite{wang2026fpagent} with its original preprocessing, features,
and decision rules, reporting mean Human-class probability and modal predicted
class. Binary comparisons across the four main configurations use two-sided
Fisher's exact tests, counting a session as Human when Human scores highest. We
map each JA4 TLS fingerprint through JA4DB to check the claimed browser
engine~\cite{foxioJA4,foxioJA4DB}, using each tool's valid-output count as its
denominator.

\noindent\textbf{AgentShield and task completion (\researchq{3}).}
We ran 30 paired trials across five seeds, two forms (\texttt{simpleForm} and
\texttt{dynamicForm}, which conditionally reveals follow-up fields and logs
reveal-to-input delay), and three configurations: \Browsium with default input,
CloakBrowser + Humanize, and \DarkAgent. A trial counts as completed only when
the server verifies that all submitted field values match expected values
exactly. Completed trials are scored by AgentShield's frozen models at a $0.5$
cutoff~\cite{agentshield}; we report completions per assigned trial, flag rates
on completed trials, and mean bot scores alongside AgentShield's released human
and default-agent logs.

\begin{table*}[t]
\centering
\caption{\textbf{Stealth Bench~V1 bypass rates (\%).}
Task counts are in parentheses. \DarkAgent was evaluated across all 80
tasks; other rows report the benchmark's historical baselines, which score
unblocked page access.}
\label{tab:stealth-bench}
\small
\setlength{\tabcolsep}{5pt}
\renewcommand{\arraystretch}{1.0}

\begin{tabularx}{\textwidth}{>{\raggedright\arraybackslash}X*{12}{r}}
\toprule
\rowcolor{black!12}
\textbf{Configuration}
  & \textbf{CF} & \textbf{PX} & \textbf{DD}
  & \textbf{Aka.} & \textbf{reC.} & \textbf{Cust.}
  & \textbf{Kas.} & \textbf{Shp.} & \textbf{GT}
  & \textbf{hC} & \textbf{Temu} & \textbf{All} \\

\rowcolor{black!12}
  & {\footnotesize(22)} & {\footnotesize(18)} & {\footnotesize(13)}
  & {\footnotesize(6)} & {\footnotesize(6)} & {\footnotesize(5)}
  & {\footnotesize(1)} & {\footnotesize(1)} & {\footnotesize(4)}
  & {\footnotesize(3)} & {\footnotesize(1)} & {\footnotesize(80)} \\

\midrule
\rowcolor{white}
Browser Use Cloud
  & 93 & 81 & 69 & 80 & 80 & 80
  & \textbf{100} & 20 & 50 & 33 & 0 & 77 \\

\rowcolor{black!5}
Anchor
  & 80 & 81 & 77 & 67 & 83 & 70
  & 0 & \textbf{100} & 38 & \textbf{67} & 0 & 74 \\

\rowcolor{white}
Onkernel
  & 80 & 31 & 62 & \textbf{100} & 89 & \textbf{100}
  & \textbf{100} & 0 & 75 & \textbf{67} & \textbf{100} & 68 \\

\rowcolor{black!5}
Browserless
  & 58 & 28 & 69 & 56 & 83 & 87
  & 0 & 0 & \textbf{83} & 56 & \textbf{100} & 57 \\

\rowcolor{white}
Local (headful)
  & 54 & 28 & 68 & 0 & 77 & 80
  & 80 & \textbf{100} & 70 & 33 & 0 & 50 \\

\rowcolor{black!5}
Steel
  & 59 & 28 & 54 & 17 & 67 & 80
  & 0 & 50 & 25 & 33 & 0 & 46 \\

\rowcolor{white}
Browserbase
  & 67 & 22 & 12 & 12 & 83 & 85
  & 50 & 0 & 50 & 50 & 0 & 43 \\

\rowcolor{black!5}
Hyperbrowser
  & 44 & 30 & 36 & 11 & 83 & 73
  & 33 & 0 & 33 & 33 & 0 & 40 \\

\rowcolor{white}
Local (headless)
  & 1 & 0 & 0 & 0 & 17 & 0
  & 33 & 0 & 0 & 11 & 0 & 2 \\

\midrule
\rowcolor{black!5}
\textbf{\DarkAgent (ours)}
  & \textbf{95} & \textbf{83} & \textbf{100}
  & \textbf{100} & \textbf{100} & 80
  & \textbf{100} & \textbf{100} & 75
  & 33 & 0 & \textbf{89} \\

\bottomrule
\end{tabularx}

\par\vspace{3pt}
\begin{minipage}{\textwidth}
\footnotesize\raggedright
CF: Cloudflare, PX: PerimeterX, DD: DataDome, Aka.: Akamai, reC.: reCAPTCHA,
Cust.: Custom, Kas.: Kasada, Shp.: Shape, GT: GeeTest, hC: hCaptcha.
\par
\end{minipage}
\end{table*}

\subsection{Behavioral Setup}
\label{sec:behavioral-setup}

\noindent\textbf{Keyboard data.}
We use the Aalto keyboard dataset of Dhakal et al.~\cite{dhakal2018typing},
which records key-down and key-up events with millisecond timestamps and target
texts. We split the data into three participant- and target-disjoint training,
validation, and test sets in approximate 72/8/20 proportions. To cover shorter
and more varied inputs, we add word-level and character-level crops during
training. Each crop keeps a contiguous substring of the target and its recorded
events and timings, provided no keys are held at either end. Keystrokes cut
from a longer sentence may not match how a person types that text on its own,
but Aalto contains only full sentences, so short fragments, typos, and unusual
punctuation are otherwise rare in training.

\noindent\textbf{Mouse data.}
We use the Bo\u{g}azi\c{c}i Mouse Dynamics Dataset~\cite{kilic2021bogazici} which records mouse trajectories. We split the data into three participant-disjoint training, validation and test sets in approximate 86/6/8 proportions. For press--release hold times, we use the Bo\u{g}azi\c{c}i Mouse Dynamics Dataset, and for scroll time, we use the Attentive Cursor Dataset~\cite{leiva2020attentive} respectively.

\section{Evaluation}
\label{sec:evaluation}
\label{sec:results}


Our evaluation addresses four research questions.
Section~\ref{sec:deployed-defenses} evaluates \DarkAgent against deployed
anti-bot defenses (\researchq{1}).
Section~\ref{sec:res-honeysite} isolates the contributions of \Browsium
and \Behavium (\researchq{2}).
Section~\ref{sec:res-rq3} examines behavioral fidelity, functional
correctness, and AgentShield form outcomes (\researchq{3}).
Finally, Section~\ref{sec:mimictrace} uses \MimicTrace to assess whether
detection based on behavioral and execution-environment evidence
generalizes to unseen devices and agent families (\researchq{4}).

\subsection{Detection Effectiveness (RQ1)}
\label{sec:deployed-defenses}

\noindent\textbf{Deployed-site blocking outcomes.}
We first evaluate whether the integrated stack passes the gates that
commercial defenses place in front of a page load. Stealth Bench~V1 answers
this with a binary access verdict per task and publishes historical baselines
for nine agent stacks~\cite{stealthbenchv1}. Running over geo-matched
residential proxies, \DarkAgent clears 71 of the 80 tasks (89\%, 95\% Wilson
CI $[80.0, 94.0]$), against 77\% for Browser Use Cloud, the strongest complete
baseline ($309/400$ across five runs on provider cloud infrastructure;
Table~\ref{tab:stealth-bench}). It clears every DataDome, Akamai, and reCAPTCHA
task and all but one Cloudflare task, and it holds the highest bypass rate,
including ties, in seven of eleven categories.
Appendix~\ref{app:evaluation-details} describes task aggregation.

\noindent\textbf{Where failures cluster.}
The nine failures concentrate on two vendors. Three of the eighteen PerimeterX
tasks and two of the three hCaptcha tasks account for more than half of them;
the rest are single blocks on Cloudflare, Custom anti-bot, GeeTest, and Temu.
A blocking outcome does not identify which channel produced it: a PerimeterX
verdict combines browser attributes, network reputation, and behavioral
scoring into one decision. The next two stages therefore separate these
channels with detectors whose inputs we can inspect.

\subsection{Component Contributions (RQ2)}
\label{sec:res-honeysite}

\noindent\textbf{Controlled detector responses.}
We next test whether a released agent detector that scores behavior can
distinguish \DarkAgent from a human, and whether the transport layer
contradicts the engine the persona claims. Table~\ref{tab:matrix} reports 18
configurations, 30 sessions each, scored concurrently by the released
\FPAgent behavioral classifier and by a JA4DB lookup of the TLS engine family.

\suppressfloats[t]

\begin{table}[t]
\centering
\caption{\textbf{\FPAgent and JA4DB results (\researchq{2}).}
Mean $P(\mathrm{Human})$ (\%), modal class, and JA4DB engine mapping across 30
sessions per stack.}
\label{tab:matrix}
\small
\setlength{\tabcolsep}{3.5pt}
\renewcommand{\arraystretch}{1.0}

\resizebox{\columnwidth}{!}{%
\begin{tabular}{lrll}
\toprule
\rowcolor{black!12}
\textbf{Configuration}
& \textbf{$P(\mathrm{Human})$}
& \textbf{Verdict}
& \textbf{TLS (JA4DB)} \\

\midrule
\rowcolor{black!9}
\multicolumn{4}{l}{
    \textit{Browser baselines (default input)}
} \\

\rowcolor{white}
Chromium & 8 & Browser Use & Chromium \\

\rowcolor{black!5}
Firefox & 14 & Manus & Firefox \\

\rowcolor{white}
Camoufox & 18 & Manus & Firefox \\

\rowcolor{black!5}
Patchright & 22 & Manus & Chromium \\

\rowcolor{white}
CloakBrowser & 0 & Manus & Chromium \\

\midrule
\rowcolor{black!9}
\multicolumn{4}{l}{
    \textit{Browser and \Behavium ablations}
} \\

\rowcolor{white}
Chromium + \Behavium & \textbf{100} & Human & Chromium \\

\rowcolor{black!5}
Chromium + \MouseGen & 100 & Human & Chromium \\

\rowcolor{white}
Chromium + \KeyboardGen & 32 & Manus & Chromium \\

\rowcolor{black!5}
\Browsium, default input & 0 & Manus & Chromium \\

\rowcolor{white}
\Browsium + \Behavium (\DarkAgent)
& \textbf{100} & Human & Chromium \\

\rowcolor{black!5}
\Browsium + \MouseGen & 98 & Human & Chromium \\

\rowcolor{white}
\Browsium + \KeyboardGen & 2 & Browser Use & Chromium \\

\midrule
\rowcolor{black!9}
\multicolumn{4}{l}{
    \textit{Public behavioral mimickers}
} \\

\rowcolor{white}
CloakBrowser + Humanize & 89 & Human & Chromium \\

\rowcolor{black!5}
Chromium + Ghost Cursor & 100 & Human & Chromium \\

\rowcolor{white}
Chromium + HumanTyping & 11 & Browser Use & Chromium \\

\rowcolor{black!5}
\Browsium + Ghost Cursor & 100 & Human & Chromium \\

\rowcolor{white}
\Browsium + HumanTyping & 9 & Browser Use & Chromium \\

\rowcolor{black!5}
\Browsium + Ghost Cursor + HumanTyping & 99 & Human & Chromium \\

\bottomrule
\end{tabular}%
}
\end{table}

Under default input, \FPAgent classifies every browser stack as automated: no
session is classified Human and mean $P(\mathrm{Human})$ stays between $0\%$
and $22\%$. Enabling \Behavium flips the verdict to Human on all 30 sessions
for both stock Chromium and \Browsium (two-sided Fisher's exact
$p = 1.69 \times 10^{-17}$). JA4DB maps all 16 Chromium-derived stacks to
Chromium and both Firefox and Camoufox to Firefox, so the transport layer
never contradicts the claimed engine.

\noindent\textbf{Mouse and keyboard contributions.}
The component ablations show how cursor activity affects this detector
in the tested configurations. \MouseGen by itself recovers almost the full effect ($100\%$ on
Chromium, $98\%$ on \Browsium), whereas \KeyboardGen by itself leaves the
verdict at automated ($32\%$ and $2\%$): changing only the typing leaves the
cursor jumping instantaneously between targets. The released 50-feature
vector counts cursor activity but omits cursor speed, acceleration, and
rollover timing~\cite{wang2026fpagent}. Public mimickers also receive high mean
Human-class probabilities: \GhostCursor ($100\%$) and CloakBrowser + Humanize
($89\%$). These results show that the tested mimickers can satisfy \FPAgent
without establishing that their traces resemble human motion. Establishing
resemblance requires inspecting the traces themselves.

\subsection{Behavioral Fidelity (RQ3)}
\label{sec:res-rq3}

We evaluate \KeyboardGen and \MouseGen on participant-disjoint human datasets
with detectors trained to separate them, and then against AgentShield's frozen
form detectors~\cite{agentshield}, which score a complete session rather than
a single trace.

\noindent\textbf{Generator. }
Across three seeds, we evaluate held-out traces using XGBoost classifier two-sample tests (C2ST) with ROC-AUC~\cite{lopezpaz2017c2st}, human-only Isolation Forest and One-Class SVM ($\mathrm{AUC}_{\pm}=\max(\mathrm{AUROC},1-\mathrm{AUROC})$), and Sliced Wasserstein Distance (SWD)~\cite{bonneel2015sliced}. We also assess zero-shot transfer (\emph{Trans.}) from detectors trained only on heuristic mimickers. No test-split data are used during training or model selection.

\noindent\textbf{Keyboard.}
We compare against HumanTyping and CloakBrowser Humanize \cite{humantyping, cloakbrowser}, which generate typing interactions directly from requested target text (Table~\ref{tab:modellevel}). Some prior methods, such as Monaco et al., Migdal and Rosenberger, and BeCAPTCHA-Type, generate timings for a supplied key sequence \cite{monaco2015spoofing,migdal2019statistical,becaptchatype}. KeyGAN \cite{keygan} generates synthetic keystroke timing features for digital phenotyping. This distinction matters because comparing these methods end to end would require either supplying part of the output that our model must generate or adding a separate event-generation policy.

We use 19 features covering both event sequences and timing gaps. Building on the Aalto typing study \cite{dhakal2018typing}, we use inter-key intervals, keypress durations, rollover ratio, and words per minute. In our evaluation, inter-key intervals and keypress durations are each summarized by their mean, standard deviation, median, and 90th percentile. However, these measures alone do not tell the whole story of how mistakes and corrections unfold. To better capture event-sequence behavior, we additionally include event efficiency, error episode rate, mean error-detection length, and mean correct run length. Table~\ref{table:keyboard-features} in Appendix~\ref{app:evaluation-details} defines these additional event features. For timing behavior, we additionally include mean correction delay and the mean, standard deviation, median, and 90th percentile of inter-event gaps. Unlike inter-key intervals, which measure the time between consecutive key-downs, or keypress durations, which measure the time between a key-down and its corresponding key-up, inter-event gaps measure the time between any two consecutive physical events. We include these features because they directly characterize the timing gaps generated by our Timing Decoder. For the feature-based classifiers, undefined values are encoded using a sentinel value of $-1$, while all defined feature values are nonnegative.

\noindent\textbf{Mouse.}
We use 11 features organized into three blocks covering spatial geometry, step timing, and movement kinematics. Building on prior mouse-dynamics and bot-detection studies~\cite{chu2013blog,kilic2021bogazici,acien2022becaptcha}, we characterize spatial path shape using log path length, path efficiency (net displacement divided by path length), root-mean-square lateral deviation from the start-to-target axis, and mean absolute turning angle across consecutive non-zero steps. Yet spatial geometry alone does not capture how cursor motion unfolds over time or how velocity changes along the path. To capture timing behavior, we include log total movement duration and the mean and standard deviation of step-level log timing gaps. To capture kinematic behavior, we include step log-speed and step log-acceleration, each summarized by its median and interquartile range. Whereas total duration summarizes overall movement time, step-level log timing gaps measure the interval preceding each spatial increment, and step acceleration measures velocity change across midpoint-spaced consecutive intervals. We include these features as they directly characterize the $(\log\Delta t_i, \Delta x_i, \Delta y_i)$ step increments generated by \MouseGen's trajectory decoder alongside their induced first- and second-order physical derivatives.

\subsubsection{Corpus Evaluation and Detector Transfer}
\label{sec:indepedent-behavioral-model-checks}
\label{sec:constraint-satisfaction}
\label{sec:architectural-justification}
\mbox{}\par\nobreak
\vspace{0.5\baselineskip}

\noindent\textbf{Behavioral resemblance and detector transfer.}
Table~\ref{tab:modellevel} shows that heuristic mimickers remain readily distinguishable from human behavior, with XGBoost ROC-AUCs of $0.9980$--$0.9999$. In contrast, \KeyboardGen and \MouseGen achieve substantially lower ROC-AUCs of $0.6809$ and $0.7197$, respectively, with one-class detectors and SWD supporting the same trend.

\begin{table}[t]
\centering
\caption{\textbf{Trace-level behavioral separability and transfer.}
Held-out ROC-AUC ($0.5$ is chance), one-class $\mathrm{AUC}_{\pm}$, SWD, and
cross-family transfer (\emph{Trans.}) over three seeds (max SD $\le 0.030$).}
\label{tab:modellevel}
\small
\setlength{\tabcolsep}{3.2pt}
\renewcommand{\arraystretch}{1.0}

\resizebox{\columnwidth}{!}{%
\begin{tabular}{lrrrrr}
\toprule
\rowcolor{black!12}
\textbf{Model}
& \textbf{XGB}
& \textbf{iForest}
& \textbf{OC-SVM}
& \textbf{SWD}
& \textbf{Trans.} \\

\midrule
\rowcolor{black!9}
\multicolumn{6}{l}{
    \textit{Keystrokes (Aalto held-out test)}
} \\

\rowcolor{white}
HumanTyping\textsuperscript{$\dagger$}
& 0.9980 & 0.8918 & 0.9946
& 3.86\textsuperscript{$\ddagger$} & 0.996 \\

\rowcolor{black!5}
CloakBrowser Humanize\textsuperscript{$\P$}
& 0.9990 & 0.7600 & 0.9400 & 1.80 & 0.972 \\

\rowcolor{white}
\textbf{\KeyboardGen (ours)}
& \textbf{0.6809} & \textbf{0.6389} & \textbf{0.6094}
& \textbf{0.183}\textsuperscript{$\ddagger$} & \textbf{0.500} \\

\midrule
\rowcolor{black!9}
\multicolumn{6}{l}{
    \textit{Mouse movement (Bo\u{g}azi\c{c}i sealed test)}
} \\

\rowcolor{white}
\GhostCursor
& 0.9999 & 0.9882 & 0.9949 & 1.993 & --- \\

\rowcolor{black!5}
CloakBrowser Humanize\textsuperscript{$\P$}
& 0.9999 & --- & --- & 2.210 & --- \\

\rowcolor{white}
\textbf{\MouseGen (ours)}
& \textbf{0.7197} & \textbf{0.5070} & \textbf{0.5224}
& \textbf{0.195} & \textbf{0.4942} \\

\bottomrule
\end{tabular}%
}

\par\vspace{3pt}
\begin{minipage}{\columnwidth}
\footnotesize\raggedright
\textsuperscript{$\dagger$}Headless HumanTyping (no keypress-duration modeling) for test columns;
real-browser \texttt{HumanTyper} (500 pairs) for \emph{Trans.}
\textsuperscript{$\ddagger$}Worst keystroke group SWD; mean of 3 blocks for mouse.
\textsuperscript{$\P$}Real-browser \texttt{cloakbrowser} v0.4.10 ($1{,}000$
mouse validation pairs, seed $42$; $500{+}497$ Aalto Split~0 pairs, also used
for keystroke \emph{Trans.}).
\par
\end{minipage}
\end{table}

Detectors trained only on heuristic mimickers generalize poorly to the learned generators, with transfer ROC-AUCs of $0.500$ for \KeyboardGen and $0.4942$ for \MouseGen. Yet detectors trained directly on their outputs recover substantial separation from human behavior. These results highlight the difficulty of developing robust detectors when representative adversarial examples are unavailable.

\noindent\textbf{Functional correctness and design choices.}
\KeyboardGen successfully reproduces $18{,}895$ of $18{,}930$ held-out targets ($99.82\%$) in a single attempt. On Split~2, shared latent conditioning and the quantized timing likelihood reduce XGBoost ROC-AUC by $7.58$ and $4.08$ percentage points, respectively, compared with their corresponding ablations. \MouseGen records no boundary violations or zero-displacement steps across $12{,}000$ generated trajectories. 

\subsubsection{AgentShield Form Evaluation}
\label{sec:res-agentshield}
\mbox{}\par\nobreak
\vspace{0.5\baselineskip}

\noindent\textbf{Session-level verdicts.}
AgentShield's two released detectors score a complete form session, which
tests whether the generators survive composition with the agent's planning
loop. Table~\ref{tab:agentshield} compares our paired runs with AgentShield's
released logs and its 11-session human baseline~\cite{agentshield}. The
detectors flag every released default agent ($78/78$ on \texttt{simpleForm},
$61/61$ on \texttt{dynamicForm}) and every \Browsium run with default input,
which emits a handful of cursor movements and sub-millisecond keypress durations.
\DarkAgent completes all five verified trials per form with zero flags on
either form (mean scores $<0.001$ and $0.019$).

\noindent\textbf{What the dynamic form exposes.}
Passing AgentShield does not by itself establish behavioral fidelity.
CloakBrowser + Humanize is also unflagged ($0/10$) because it emits high
cursor-movement volume, even though its mean keypress durations are nearly four times
shorter than those of humans and \DarkAgent (Table~\ref{tab:agentshield}).
The conditional form does expose planning latency: on \texttt{dynamicForm},
\DarkAgent reacts to a newly revealed field in $3.74$\,s on average
($3.28$--$4.17$\,s), faster and more regular than humans at $4.72$\,s
($2.15$--$9.02$\,s), with CloakBrowser + Humanize at $3.80$\,s and \Browsium
with default input at $0.03$\,s. These results show that passing AgentShield
does not eliminate differences in the timing between actions.

\begin{table}[t]
\centering
\caption{\textbf{AgentShield form evaluation.}
Verified completions ($c/N$), flags at the $0.5$ cutoff ($k/c$), mean bot
score, cursor movements, and keypress duration (ms) across both released
detectors~\cite{agentshield}. \DarkAgent includes the pre-click pause policy
ablated in Table~\ref{tab:mimictrace-summary}.}
\label{tab:agentshield}
\small
\setlength{\tabcolsep}{3.0pt}
\renewcommand{\arraystretch}{1.0}

\resizebox{\columnwidth}{!}{%
\begin{tabular}{llrrrrr}
\toprule
\rowcolor{black!12}
\textbf{Form}
& \textbf{Configuration}
& \textbf{Comp.}
& \textbf{Flag}
& \textbf{Score}
& \textbf{Moves}
& \shortstack{\textbf{Keypress}\\\textbf{duration}} \\

\midrule
\rowcolor{black!9}
\multicolumn{7}{l}{
    \textit{\texttt{simpleForm}}
} \\

\rowcolor{white}
& Human reference (released)
& 5/5 & 0/5 & 0.027 & 110.2 & 114.8 \\

\rowcolor{black!5}
& Default agents (8 stacks)
& 78/78 & 78/78 & 0.973 & 20.6 & --- \\

\rowcolor{white}
& \Browsium, default input
& 5/5 & 5/5 & 0.980 & 4.6 & 0.1 \\

\rowcolor{black!5}
& CloakBrowser + Humanize
& 5/5 & 0/5 & $<$0.001 & 278.0 & 30.3 \\

\rowcolor{white}
& \textbf{\DarkAgent (\Browsium + \Behavium)}
& \textbf{5/5} & \textbf{0/5} & \textbf{$<$0.001}
& 178.2 & 121.3 \\

\midrule
\rowcolor{black!9}
\multicolumn{7}{l}{
    \textit{\texttt{dynamicForm}}
} \\

\rowcolor{white}
& Human reference (released)
& 6/6 & 0/6 & 0.040 & 174.7 & 107.7 \\

\rowcolor{black!5}
& Default agents (6 stacks)
& 61/61 & 61/61 & 0.960 & 21.6 & --- \\

\rowcolor{white}
& \Browsium, default input
& 5/5 & 5/5 & 0.562 & 4.0 & 0.2 \\

\rowcolor{black!5}
& CloakBrowser + Humanize
& 5/5 & 0/5 & 0.004 & 190.0 & 24.7 \\

\rowcolor{white}
& \textbf{\DarkAgent (\Browsium + \Behavium)}
& \textbf{5/5} & \textbf{0/5} & \textbf{0.019}
& 162.2 & 127.6 \\

\bottomrule
\end{tabular}%
}
\end{table}

\subsection{Signal Migration}
\label{sec:defensive-implications}

Across \researchq{1}--\researchq{3}, \DarkAgent passed the tested transport and session checks, yet its timing between actions still differed from human timing.
Section~\ref{sec:mimictrace} tests whether combining signals from the same session supports detection on unseen devices and agent families.

\section{Multi-Channel Detection and Generalization (RQ4)}
\label{sec:mimictrace}

To address  \researchq{4}, we introduce \MimicTrace, an open benchmark that links
behavioral and execution-environment telemetry with server-verified task
outcomes. We use it to examine whether signals that distinguish agent
and human sessions within an evaluated cohort also support detection
on unseen devices and agent families.

\subsection{Benchmark Design}
\label{sec:mimictrace-design}

\begin{figure}[t]
  \centering
  \includegraphics[width=0.82\columnwidth]
    {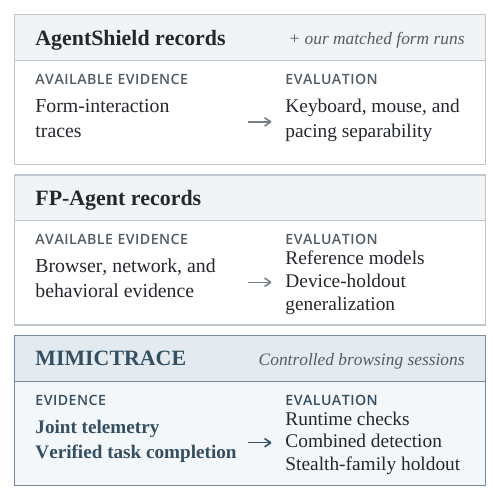}
  \caption{\textbf{\MimicTrace overview.}
  Controlled \MimicTrace
sessions complement imported records with joint telemetry and
verified task completion.}
\label{fig:mimictrace-data-sources}
\end{figure}

The evaluations in Section~\ref{sec:evaluation} provide complementary
measurements. Stealth Bench~V1 reports live-site access outcomes.
\FPAgent records browser fingerprints and behavioral statistics, but
its classification verdict does not establish task completion.
AgentShield evaluates form interactions without recording network or
browser telemetry. \MimicTrace supports joint analysis of network,
browser, control, and behavioral evidence
(Figure~\ref{fig:mimictrace-data-sources}), with three capabilities.

\noindent\textbf{Joint telemetry with verified task outcomes.}
The benchmark hosts browsing and form tasks and verifies submitted values
on the server. Each newly collected session links network headers,
browser properties, control and runtime probes, and raw mouse and
keyboard timestamps to its task outcome in a common record. Imported
logs retain their available fields, with missing channels represented
explicitly.

\noindent\textbf{Frozen defenses and diagnostic classifiers.}
Frozen defenses retain their original models and decision rules to
measure acceptance by existing detectors. Diagnostic classifiers are
trained on recorded features and evaluated on held-out data to test
for residual differences between agent and human sessions. These results
characterize separability within the evaluated cohorts; separate transfer
experiments test generalization across devices and agent families.

\noindent\textbf{Consistency checks and environment models.}
Learned fingerprint models can respond to host characteristics absent
from their human reference. \MimicTrace therefore pairs these models
with training-free consistency checks~\cite{vastel2018fpscanner,venugopalan2024fpinconsistent}, such as agreement between the
platform reported by HTTP headers and JavaScript APIs. Reporting rule
violations separately from model predictions helps distinguish explicit
inconsistencies from detections that depend on the reference cohort.

\subsection{Evaluation Protocol}
\label{sec:mimictrace-schema}
\label{sec:mimictrace-cohorts}

\noindent\textbf{Benchmark tasks and session record.}
\MimicTrace serves three pages on a local web server: an interactive browsing
page, a static form, and a conditional form that reveals follow-up fields after
earlier inputs. Each visit writes one JSON record with four sections. The
network section logs the TLS JA4 fingerprint and HTTP header order while
stripping IP and location data. The browser section logs \texttt{navigator},
screen, graphics, and audio properties. The control and runtime section
compares values across the main page, an \texttt{iframe}, and a web worker, and
checks browser input rules such as click pressure and key-event order. The
behavioral section logs raw mouse, scroll, and keyboard events with millisecond
timestamps.

\noindent\textbf{Combining public logs with new visits.}
To let defenders compare against prior baselines or bring their own traces,
\MimicTrace also imports public logs from AgentShield and \FPAgent into the
same schema. When importing an outside log, we compute extra features only from
events that log actually recorded and leave missing sections blank rather than
filling them in. For forms (190 sessions), we pair AgentShield's 150 public
logs (11 human sessions from six authors and 139 default-agent runs, containing
form events only) with 40 new server-checked runs on the same forms. General
browsing and standalone typing or mouse datasets are kept out of this human
baseline so every trace comes from the same task. For browsing environments, we
pair \FPAgent's $7{,}719$ public records (546 human and $7{,}173$ agent) with a
172-session honeysite audit: 108 newly recorded four-channel visits covering
stock Chromium, headful nodriver Chrome, and \Browsium with default input or
\Behavium, a descriptor-emulated nodriver (emul.\ CDP) + \Behavium control, and
single-generator \Browsium ablations, plus 64 imported visits across six
external stealth tools that carry network, browser, and behavioral telemetry
without runtime payloads. We also record 14 normal desktop browser launches to
verify our runtime rules (Table~\ref{tab:mimictrace-channels};
Appendix~\ref{app:mimictrace-details}).

\begin{table*}[t]
\centering
\caption{\textbf{\MimicTrace form C2ST.}
Frozen AgentShield flag rate ($k/N$), per-channel XGBoost C2ST ROC-AUC
(mean $\pm$ SD of pooled out-of-fold ROC-AUC across 20 repeats of stratified
5-fold CV vs.\ 11 human sessions), and median pre-click pause.}
\label{tab:mimictrace-summary}
\small
\setlength{\tabcolsep}{5pt}
\renewcommand{\arraystretch}{1.0}

\begin{tabularx}{\textwidth}{>{\raggedright\arraybackslash}Xrrrrr}
\toprule
\rowcolor{black!12}
\textbf{Configuration}
& \textbf{Flagged}
& \textbf{Keyboard}
& \textbf{Mouse}
& \textbf{Pacing}
& \textbf{Pause (ms)} \\

\midrule
\rowcolor{black!9}
\multicolumn{6}{l}{
    \textit{Human and baseline cohorts}
} \\

\rowcolor{white}
Human reference\textsuperscript{$\dagger$}
  & 0/11 & 0.510 & 0.486 & 0.488 & 58.0 \\

\rowcolor{black!5}
Default agents (8 stacks, 14 form entries)
  & 139/139 & 1.000 $\pm$ 0.000 & 0.996 $\pm$ 0.007
  & 0.949 $\pm$ 0.016 & 2.8 \\

\rowcolor{white}
\Browsium, default input
  & 10/10 & 1.000 $\pm$ 0.000 & 1.000 $\pm$ 0.000
  & 1.000 $\pm$ 0.000 & 10.5 \\

\rowcolor{black!5}
CloakBrowser + Humanize
  & 0/10 & 1.000 $\pm$ 0.000 & 0.990 $\pm$ 0.021
  & 0.739 $\pm$ 0.052 & 162.5 \\

\midrule
\rowcolor{black!9}
\multicolumn{6}{l}{
    \textit{\DarkAgent live \Browsium ablations (5 seeds $\times$ 2 forms)}
} \\

\rowcolor{white}
\DarkAgent (no pre-click pause)
  & 0/10 & 0.680 $\pm$ 0.105 & 0.986 $\pm$ 0.030
  & 0.687 $\pm$ 0.070 & 15.5 \\

\rowcolor{black!5}
\textbf{\DarkAgent (+ pre-click pause)}\textsuperscript{$\ddagger$}
  & \textbf{0/10} & \textbf{0.646 $\pm$ 0.058}
  & \textbf{0.782 $\pm$ 0.079} & \textbf{0.594 $\pm$ 0.093}
  & \textbf{63.5} \\

\bottomrule
\end{tabularx}

\par\vspace{3pt}
\begin{minipage}{\textwidth}
\footnotesize\raggedright
\textsuperscript{$\dagger$}200-shuffle permutation-null mean.
\textsuperscript{$\ddagger$}Pre-click pauses sampled from 43 browser-disjoint
\FPAgent human groups; Keyboard $p{=}0.040$ and Mouse $p{=}0.015$ vs.\
CloakBrowser + Humanize.
\par
\end{minipage}
\end{table*}

\noindent\textbf{Three types of detection tests.}
Across Section~\ref{sec:mimictrace}, we keep three kinds of instruments
distinct so their roles stay clear. \emph{Existing frozen detectors}
(AgentShield and \FPAgent) run unchanged to test whether released defenses flag
a visit. \emph{Trained classifiers} include our diagnostic XGBoost models on
form visits (Section~\ref{sec:mimictrace-evaluation}) and our \FPAgent-trained
environment and behavioral models transferred to the honeysite
(Sections~\ref{sec:mimictrace-channels}--\ref{sec:informed-defender}).
\emph{Rule-based consistency checks} require no training. We distinguish
\emph{cross-channel checks} for agreement across reported browser and network
properties from \emph{runtime checks} for engine, execution-context, and input-event
behavior. Each check tests an expected relationship, or invariant.

\subsection{Behavioral Diagnostics}
\label{sec:mimictrace-evaluation}

Section~\ref{sec:indepedent-behavioral-model-checks} evaluates individual traces
from standalone human datasets, while Section~\ref{sec:res-agentshield} tests
live form sessions against AgentShield's frozen detectors. Here, we examine
whether trained diagnostic classifiers can distinguish the completed form
sessions from human sessions, even when AgentShield does not flag them.
We use diagnostic XGBoost classifier two-sample tests
(C2STs)~\cite{lopezpaz2017c2st}. Frozen AgentShield stays unchanged and tests
whether the released defense flags a visit. The C2STs are instead trained on
the completed visits to separate each agent's keyboard, mouse, or pacing
features from the 11 human form sessions. Each score averages 20 repeats of
five-fold cross-validation and is compared against 200 random label shuffles.
Table~\ref{tab:mimictrace-summary} shows why both tests matter: CloakBrowser +
Humanize and \DarkAgent both pass frozen AgentShield with $0/10$ flags, yet
C2ST separates Humanize almost completely in keyboard ($1.000$) and mouse
($0.990$) views while \DarkAgent lowers keyboard separation to $0.680$ and
$0.646$ ($p=0.040$ vs.\ Humanize).

\noindent\textbf{Why mouse separation stays high before clicking.}
\DarkAgent's first row in Table~\ref{tab:mimictrace-summary} presents a clear
puzzle: live \MouseGen matches humans in movement speed
($\mathrm{AUC}_{\pm}=0.500$) and path straightness ($0.527$), yet full mouse
C2ST still reaches $0.986$. The gap lies at the seam between moving and
clicking. \MouseGen moves the cursor onto the target, after which the browser
presses the mouse button. Humans pause briefly on the target before pressing
($58.0$\,ms median from the last \texttt{mousemove} to \texttt{mousedown}),
whereas pressing right away leaves a median \emph{pre-click pause} of only
$15.5$\,ms. To test this seam directly, we keep \Browsium, \KeyboardGen, and
\MouseGen fixed and sample pre-click pauses from 43 separate \FPAgent human
browser groups, using no AgentShield human data. Across the same five seeds and
two forms, the median pause moves to $63.5$\,ms and mouse C2ST drops from
$0.986$ to $0.782$ ($p=0.015$ vs.\ Humanize), while all ten runs finish with
$0/10$ AgentShield flags.

\KeyboardGen itself does not change between the two \DarkAgent rows, so the
shift in keyboard C2ST ($0.680$ vs.\ $0.646$) cannot be attributed to a
change in the keyboard model; these separate live runs do not isolate its cause. Adding
human pauses narrows the mouse gap, though mouse C2ST stays above the
shuffled-label baseline ($p=0.030$). Pacing C2ST ($0.594$) does not differ
significantly from that chance baseline ($p=0.264$). The small sample
limits interpretation; nonsignificance does not establish human equivalence. Together, these runs show
how diagnostic C2STs pinpoint timing leaks between sub-actions that frozen
session detectors miss.

\begin{table*}[t]
\centering
\caption{\textbf{Rule-based audit and combined detection.} \emph{CC} and
\emph{RI} count sessions flagged by cross-channel checks ($0/546$ \FPAgent humans)
and runtime checks ($0/14$ stock controls). \emph{Ref.} and \emph{Beh.}
report cohort-relative digest novelty ($0.0\%$ FPR) and the behavioral random
forest ($2.58\%$ FPR); \emph{OR} flags their union ($2.58\%$ FPR on $155$
held-out \FPAgent humans, which does not estimate a local honeysite human rate).}
\label{tab:mimictrace-channels}
\footnotesize
\setlength{\tabcolsep}{3.5pt}
\renewcommand{\arraystretch}{1.0}
\begin{tabularx}{\textwidth}{>{\raggedright\arraybackslash}Xrrrrrr
>{\raggedright\arraybackslash}p{0.26\textwidth}}
\toprule
\rowcolor{black!12}
\textbf{Configuration} & $\boldsymbol{n}$ & \textbf{CC} & \textbf{RI}
& \textbf{Ref.} & \textbf{Beh.} & \textbf{OR} & \textbf{Primary signal} \\
\midrule
\rowcolor{black!9}
\multicolumn{8}{l}{\textit{Controlled $3\times 2$ browser $\times$ behavior comparison, descriptor control, and \Browsium ablations}} \\
\rowcolor{white}
Chromium, default input & $12$ & $12/12$ & $0/12$ & $0/12$ & $12/12$ & $12/12$ & Unreduced UA, no \texttt{chrome}, behavior \\
\rowcolor{black!5}
Chromium + \Behavium & $12$ & $12/12$ & $0/12$ & $0/12$ & $0/12$ & $12/12$ & Unreduced UA, no \texttt{chrome}, WebGL \\
\rowcolor{white}
Headful nodriver, default input & $12$ & $0/12$ & $12/12$ & $0/12$ & $12/12$ & $12/12$ & Zero-pressure press, \texttt{char}-only keys \\
\rowcolor{black!5}
Headful nodriver + \Behavium & $12$ & $0/12$ & $12/12$ & $0/12$ & $4/12$ & $12/12$ & Zero-pressure CDP \texttt{pointerdown} \\
\rowcolor{white}
Descriptor-emulated nodriver (emul.\ CDP) + \Behavium & $12$ & $0/12$ & $0/12$ & $0/12$ & $3/12$ & $3/12$ & CDP frame-coalesced cursor density \\
\rowcolor{black!5}
\Browsium, default input & $12$ & $0/12$ & $0/12$ & $0/12$ & $12/12$ & $12/12$ & Behavioral random forest \\
\rowcolor{white}
\textbf{\Browsium + \Behavium (\DarkAgent)} & $\mathbf{12}$ & $\mathbf{0/12}$ & $\mathbf{0/12}$ & $\mathbf{0/12}$ & $\mathbf{0/12}$ & $\mathbf{0/12}$ & \textbf{None detected} \\
\rowcolor{black!5}
\Browsium + \MouseGen / +\KeyboardGen & $24$ & $0/24$ & $0/24$ & $0/24$ & $24/24$ & $24/24$ & Unmodeled keyboard or mouse \\
\midrule
\rowcolor{black!9}
\multicolumn{8}{l}{\textit{External stealth baselines (zero-shot transfer)}} \\
\rowcolor{white}
CloakBrowser + Humanize & $12$ & $0/12$ & --- & $12/12$ & $0/12$ & $12/12$ & Audio/canvas digest novelty \\
\rowcolor{black!5}
Camoufox & $15$ & $12/15$ & --- & $0/15$ & $15/15$ & $15/15$ & OS fonts, locale, behavior \\
\rowcolor{white}
Patchright & $12$ & $12/12$ & --- & $0/12$ & $12/12$ & $12/12$ & Missing \texttt{chrome}, UA, behavior \\
\rowcolor{black!5}
Headless nodriver / uc-chrome / BU & $25$ & $0/25$ & --- & $0/25$ & $24/25$ & $24/25$ & \texttt{HeadlessChrome} signal, behavior \\
\midrule
\rowcolor{black!9}
\multicolumn{8}{l}{\textit{\FPAgent reference rule audit}} \\
\rowcolor{white}
Humans & $546$ & $0/546$ & --- & --- & --- & --- & No cross-channel violations \\
\rowcolor{black!5}
ChatGPT Agent / Manus & $1{,}983$ & $1{,}954/1{,}983$ & --- & --- & --- & --- & OS fonts, UA/client-hint mismatch \\
\rowcolor{white}
Five other agents & $5{,}190$ & $1/5{,}190$ & --- & --- & --- & --- & One Browser Use mismatch \\
\bottomrule
\end{tabularx}
\par\vspace{2pt}
\begin{minipage}{\textwidth}
\footnotesize\raggedright
Headful nodriver drives stock Chrome~157 over CDP (\texttt{navigator.webdriver}
false); descriptor-emulated nodriver (emul.\ CDP) supplies WebDriver-grade force ($0.5$) and virtual
key codes. Single-generator \Browsium ablations comprise $12$ \MouseGen-only
and $12$ \KeyboardGen-only runs. Without \emph{Ref.}
($\emph{CC}\lor\emph{RI}\lor\emph{Beh.}$), controlled and ablation counts are
unchanged and $138/172$ sessions are flagged ($150/172$ with \emph{Ref.}).
Final baseline row: headless nodriver ($12$), uc-chrome ($12$), Browser Use
($1$); dashes denote imported traces without runtime payloads.\par
\end{minipage}
\end{table*}

\subsection{Environment Generalization}
\label{sec:mimictrace-channels}
The audit examines HTTP headers, browser-surface attributes, cross-channel
consistency, and runtime behavior. It excludes IP, ASN, and geolocation
data and filters collection-side header metadata. Runtime probes compare
engine capabilities, native builtins, observations across page, iframe,
and worker contexts, and Pointer/UI Events input properties.
Appendix~\ref{app:mimictrace-details} specifies the feature groups,
filtering controls, and audited browser configuration.

\noindent\textbf{Device holdout changes the generalization result.}
On \FPAgent, logistic regression reaches $0.980$ ROC-AUC under five-fold
browser-group-disjoint cross-validation. However, its seven agent families
occupy only ten browser groups and seven devices. Holding out entire devices
lowers ROC-AUC to $0.618$; browser-surface features alone score $0.540$.
Cross-channel consistency retains more separation ($0.774$).
Browser-group separation therefore does not ensure generalization to
unseen execution environments.

\noindent\textbf{Reference novelty and consistency measure different properties.}
Detectors fitted on 391 \FPAgent training humans are used on 172
honeysite sessions, with operating points capped at $5\%$ FPR on 155
device-disjoint \FPAgent calibration humans. Browser-surface XGBoost yields
$0.245$--$0.407$ ROC-AUC across honeysite stacks, showing poor transfer.
Cross-channel and runtime checks flag stock Chromium, default-CDP
headful nodriver, and several stealth baselines, but none of the 48
\Browsium sessions, 546 \FPAgent humans, or 14 stock-browser controls
(Table~\ref{tab:mimictrace-channels}).

CloakBrowser + Humanize triggers no cross-channel check, yet the
reference-augmented model flags all 12 sessions. Its audio and canvas
digests are absent from the human reference. Such novelty is relative to
the reference cohort and does not establish that these digests cannot
occur on human devices.

\subsection{Combined Detection and Family Holdout}
\label{sec:informed-defender}
We flag a session if any of three components flags it: the reference-augmented
environment model, a behavioral random forest, or the rule-based consistency checks.
This OR rule detects $150/172$ honeysite sessions ($138/172$ without reference
novelty), including every uncoordinated cell of the $3\times 2$ comparison and
all 36 partial \Browsium ablations, $3/12$ descriptor-emulated nodriver
(emul.\ CDP) + \Behavium sessions, and none of the 12 full \DarkAgent sessions
(Table~\ref{tab:mimictrace-channels}). Its $2.58\%$ human FPR is measured on
held-out \FPAgent humans, not local honeysite users.

Three-seed leave-one-stealth-family-out evaluation against unseen human
device groups yields $0/12$ \DarkAgent detections and $0.532$ ROC-AUC with
\Browsium held out. Other held-out families remain more separable
(Appendix~\ref{app:mimictrace-transfer}). Test FPR varies from $3.2\%$ to
$7.9\%$, so these detection rates are not comparable at a common observed FPR.

\section{Discussion and Limitations}
\label{sec:discussion}

\noindent\textbf{Human and platform scope.}
Our human baselines are small or external to the local testbed. Form diagnostics use 11 sessions completed by six authors of the AgentShield study, and offline mouse tests cover three held-out Bo\u{g}azi\c{c}i users. Honeysite false-positive rates ($2.58\%$) are calibrated on held-out \FPAgent desktop records rather than local visitors across macOS, WebKit, mobile browsers, privacy extensions, or assistive tools.

\noindent\textbf{System and deployment bounds.}
We evaluated \Browsium as an integrated Windows/Chromium stack rather than ablating individual browser patches. Stealth Bench~V1 baselines also reflect released historical provider runs rather than contemporaneous tests from identical residential networks. Observing $0/12$ flags on \DarkAgent honeysite sessions therefore bounds the tested passive checks without establishing resistance to unseen probes.

\noindent\textbf{Future directions.}
\Behavium generates movement and typing within single actions, leaving reading pauses and delays between actions open. Extending \MimicTrace with local human visits across operating systems, extensions, and longitudinal re-runs would measure false positives directly as browsers and defenses change. Finally, current agents receive a page observation, plan, and then act on a fixed target. Subtle interface changes during an ongoing action may expose that delay, provided false positives under network lag and assistive input remain low.
\section{Conclusion}
\label{sec:conclusion}
We presented \DarkAgent, a web-agent architecture that treats browser
execution, network identity, and interaction behavior as coupled
sources of detectability. \Browsium separates privileged observation
from native input delivery and coordinates browser and network
identity, while \MouseGen and \KeyboardGen combine learned interaction
generation with explicit physical and task constraints. Our evaluation
demonstrates successful execution under the tested defenses and shows
why detector acceptance alone cannot characterize the observable
differences between agents and humans. We also introduced \MimicTrace,
a multi-channel benchmark that combines joint
telemetry with server-verified task completion, enabling analysis of
individual signals and cross-channel relationships. Together, these
contributions connect the design of coordinated agents with systematic
measurement of the evidence their execution exposes. The findings
motivate joint analysis of detection signals across the complete
session, with task utility and residual detectability assessed
separately. A central challenge for future defenses is to identify
cross-channel inconsistencies that remain informative as agents
improve, while accounting for legitimate variation across users,
devices, and execution environments.

\clearpage
\bibliographystyle{IEEEtran}
\bibliography{references}   

\appendices

\section{Browser Background}
\label{sec:background}

Understanding coordinated evasion requires examining the topology of an
automated browsing session: how an external controller connects to the browser,
how Chromium isolates processes and scripts internally, and where a website
observes client signals.

\subsection{Controller-to-Browser Boundary}
\label{sec:bg-control}

An autonomous web agent runs outside the browser and needs a control path to
read page state and trigger actions. Frameworks such as Playwright and Puppeteer
use the Chrome DevTools Protocol (CDP) to control Chromium. CDP can use a local
debugging port or a pipe; a pipe avoids opening a debugging port but still
uses CDP inspector sessions for renderer observation and execution.
Depending on the configuration, automation can expose three classes of artifacts:
\begin{itemize}[leftmargin=*]
  \item \textbf{Debugging sockets and WebDriver state.} Chromium can expose a
  localhost port for remote debugging. Separately, the WebDriver standard
  requires \texttt{navigator.webdriver = true} during a WebDriver
  session~\cite{webdriver}. Defender scripts inside
  the page can read \texttt{navigator.webdriver} directly or probe local ports
  to detect an external controller.
  \item \textbf{Renderer inspector attachment.} To inspect DOM nodes or run
  expressions, CDP enables inspector domains such as \texttt{Runtime.enable}
  inside the renderer's JavaScript engine. Attaching this inspector emits
  execution-context events and triggers object getters during serialization,
  producing side effects that page scripts can provoke and observe.
  \item \textbf{Main-world JavaScript patches.} Stealth plugins try to hide
  these automation artifacts by injecting scripts into the page before website
  code runs to overwrite built-in properties. Because injected scripts and
  defender scripts share the same JavaScript environment, defenders can call
  \texttt{Object.getOwnPropertyDescriptor} or
  \texttt{Function.prototype.toString} to see that native browser getters were
  replaced by wrapper functions.
\end{itemize}

\subsection{Internal Browser Boundaries}
\label{sec:bg-isolation}

Internally, Chromium splits privileges across operating-system processes
connected by \emph{Mojo}, its internal inter-process communication (IPC)
message bus. One privileged \emph{browser process} owns windows, networking,
and operating-system input, while multiple sandboxed \emph{renderer processes}
isolate untrusted web pages. Inside each renderer process runs \emph{Blink},
Chromium's rendering engine, which parses HTML and CSS into the Document Object
Model (DOM) tree, computes visual layout, and exposes native DOM objects and
web APIs to JavaScript through a binding layer. Within each frame, Blink
further partitions JavaScript execution into isolated environments called
\emph{worlds} that share one underlying DOM tree: website and defender scripts
run in the \emph{main world}, whereas browser extensions run in \emph{isolated
worlds} with separate global objects and prototype chains that main-world
scripts cannot read.

Four internal boundaries govern how an agent can observe a page and deliver
input across these processes and worlds:
\begin{itemize}[leftmargin=*]
  \item \textbf{Closed shadow roots.} Web components can encapsulate internal
  DOM nodes inside a shadow tree created with \texttt{mode:\,'closed'}, for
  which \texttt{Element.shadowRoot} returns \texttt{null} across Blink's
  binding layer in both main and isolated worlds. Anti-bot widgets place
  interactive controls inside closed shadow roots to block DOM queries, and
  also create test closed roots in the main world to catch stealthy browsers
  that globally expose them.
  \item \textbf{Cross-origin out-of-process iframes (OOPIFs).} For site
  isolation, Chromium renders cross-origin subframes, such as embedded
  challenge widgets, in separate renderer processes with frame-local coordinate
  systems. Scripts in the parent frame cannot inspect an OOPIF's DOM or
  convert its local bounds into top-level viewport coordinates without
  browser-process coordination over Mojo.
\item \textbf{Top-down input path and \texttt{Event.isTrusted}.} Physical
mouse, keyboard, and scroll events enter the browser process from the
operating system, pass down through compositor coordinate transformation and
cross-frame hit-testing, and arrive at DOM listeners with the read-only flag
\texttt{Event.isTrusted = true}. Events created inside JavaScript via
\texttt{dispatchEvent} or \texttt{element.click()} originate inside the
renderer, bypass hit-testing, and carry \texttt{isTrusted = false}.
Browser-level CDP input commands do set \texttt{isTrusted = true}, but
require an active DevTools session and can leave default-omitted event
fields (such as zero pointer pressure or missing key codes) exposed to
page-level checks.
  \item \textbf{Window focus state.} The browser process tracks whether a
  window holds active operating-system focus and exposes this state via
  \texttt{document.hasFocus()} and \texttt{focus}/\texttt{blur} events.
  Headless or backgrounded browsers often lack active focus, allowing defenses
  to flag interactions that arrive while \texttt{document.hasFocus()} is
  \texttt{false}.
\end{itemize}

\subsection{Client-to-Website Observation Channels}
\label{sec:bg-channels}

At the outer boundary, a defender observes the client from two vantage points:
at the remote server receiving network traffic, and inside the page's main
world where collector scripts query browser APIs and listen to DOM events.
Alongside the \textbf{automation-control channel} (Appendix~\ref{sec:bg-control})
and the \textbf{behavioral channel} (timestamped mouse paths, speed and
acceleration, click hold durations, scroll cadence, and key-down/key-up
timings), defenses cross-check detection signals between the \textbf{network}
and \textbf{browser} channels:

\begin{itemize}[leftmargin=*]
  \item \textbf{TLS handshake and HTTP headers.} When opening an HTTPS
  connection, the browser sends an unencrypted TLS \texttt{ClientHello} message
  listing supported cipher suites, extensions, and signature algorithms
  (summarized by fingerprints such as JA4~\cite{foxioJA4}), followed by ordered
  HTTP headers including \texttt{User-Agent} and \texttt{Sec-CH-UA*} Client
  Hints. If a client changes \texttt{navigator.userAgent} only in JavaScript,
  the unchanged TLS handshake and HTTP headers on the wire expose the
  contradiction.
  \item \textbf{WebRTC network paths and media devices.} Web Real-Time
  Communication (\texttt{RTCPeerConnection}) discovers peer-to-peer routes by
  sending UDP packets to STUN servers, which echo back the client's public IP
  address. Because HTTP proxies do not tunnel raw UDP traffic by default, those
  probes exit through the machine's direct network interface and reveal the
  host's real IP address alongside the proxy IP. Its companion API,
  \texttt{navigator.mediaDevices.\allowbreak enumerateDevices()}, lists microphones,
  speakers, and cameras; cloud servers without audio hardware typically return
  an empty list, whereas consumer desktops report default audio inputs and
  outputs.
  \item \textbf{Client identity and hardware APIs.} Inside the main world,
  \texttt{navigator.userAgent}, \texttt{navigator.platform},
  \texttt{navigator.languages}, and \texttt{navigator.userAgentData} report the
  browser version, operating system, and locale, while
  \texttt{navigator.hardwareConcurrency} and \texttt{navigator.deviceMemory}
  report CPU core count and approximate RAM in gigabytes. Defenses flag
  automation substrings such as \texttt{"HeadlessChrome"}, single-language
  lists typical of unconfigured containers, or implausible hardware pairings
  such as 4 CPU cores with 64\,GB of RAM.
  \item \textbf{Display geometry and installed fonts.} \texttt{window.screen}
  reports monitor resolution, while \texttt{window.innerHeight} and
  \texttt{window.outerHeight} report the page viewport height and the full
  window height including the tab bar and toolbar. On a desktop browser
  \texttt{outerHeight} exceeds \texttt{innerHeight} by the toolbar height,
  whereas headless browsers often report equal heights because no window frame
  exists. Pages also measure rendered text widths across CSS font names to
  infer installed fonts, flagging clients that claim Windows while resolving
  Linux or macOS system fonts.
  \item \textbf{Graphics and audio pipelines.} 2D Canvas and WebGL render
  graphics and expose GPU vendor and renderer names, hardware parameter limits,
  and pixel readbacks, while WebAudio's \texttt{AudioContext} reports sample
  rate and output latency. Server environments without consumer GPUs or sound
  cards expose software-renderer strings or GPU and audio parameters that
  disagree with the claimed operating system.
\end{itemize}

\section{Supplementary Evaluation Details}
\label{app:evaluation-details}

\noindent\textbf{Stealth Bench~V1 aggregation.}
All rows in Table~\ref{tab:stealth-bench}, including historical provider
baselines, are computed directly from the released 80-task run logs rather than
the 72-task subset shown in the benchmark's summary plots (which omit GeeTest,
hCaptcha, and Temu)~\cite{stealthbenchv1,stealthbenchv1repository}. Every
configuration in Table~\ref{tab:stealth-bench} therefore shares the same
80-task denominator.

\noindent\textbf{Baseline availability.}
We requested the trajectory-generation code and benchmark data for
BeCAPTCHA-Mouse~\cite{acien2022becaptcha} from its authors on October~2 and
October~9, 2026, and had received no response as of November~2, 2026. For
DMTG~\cite{liu2024dmtg}, we could not locate a public implementation or
reproduce the generator from the available documentation. We therefore do
not report direct experimental comparisons with these generators.

\noindent\textbf{Keyboard event features.}
Table~\ref{table:keyboard-features} defines the sequence-level keyboard
features introduced in Section~\ref{sec:res-rq3}.
\begin{table}[t]
\centering
\caption{\textbf{Additional sequence-level keyboard features.}}
\label{table:keyboard-features}
\small
\setlength{\tabcolsep}{4pt}
\renewcommand{\arraystretch}{1.12}
\begin{tabularx}{\columnwidth}{>{\raggedright\arraybackslash}p{0.38\columnwidth}>{\raggedright\arraybackslash}X}
\toprule
\rowcolor{black!12}
\textbf{Feature} & \textbf{Definition} \\
\midrule
\rowcolor{white}
Event efficiency & Minimum required events divided by observed events \\
\rowcolor{black!5}
Error episode rate & Error episodes divided by target character length \\
\rowcolor{white}
Mean error-detection length & Characters typed after an error before correction begins \\
\rowcolor{black!5}
Mean correct run length & Correct character insertions between consecutive error episodes \\
\rowcolor{white}
Rollover ratio & Key-downs overlapping the preceding key hold divided by total key-downs \\
\bottomrule
\end{tabularx}
\end{table}

\section{Supplementary \MimicTrace Details}
\label{app:mimictrace-details}

\subsection{Schema Redaction, Cohorts, and Splits}
\label{app:mimictrace-cohorts}

The local server mounts \texttt{/browsing}, \texttt{/forms/simple}, and
\texttt{/forms/dynamic} and writes JSONL records under schema
\texttt{mimictrace.session.v1}. Packaging redacts \texttt{Authorization},
\texttt{Cookie}, \texttt{Set-Cookie}, and forwarding headers while preserving
wire header order, and removes IP addresses and vendor identifiers. Imported
external logs populate only their recorded channels; missing fields remain
null rather than imputed.

\begin{table*}[t]
\centering
\caption{\textbf{Cohorts and experimental roles.} Archive composition is
distinguished from protocol-specific samples; counts are reported for each
role rather than as a sum of distinct sessions.}
\label{tab:mimictrace-cohorts}
\small
\setlength{\tabcolsep}{5pt}
\renewcommand{\arraystretch}{1.0}
\begin{tabularx}{\textwidth}{>{\raggedright\arraybackslash}p{0.21\textwidth}
>{\raggedright\arraybackslash}p{0.13\textwidth}
>{\raggedright\arraybackslash}X}
\toprule
\rowcolor{black!12}
\textbf{Cohort} & \textbf{Size} & \textbf{Composition and use} \\
\midrule
\rowcolor{white}
Archived browsing & 510 sessions & 17 of 18 Table~\ref{tab:matrix} configurations, 30 sessions each (unhumanized CloakBrowser omitted); browsing portion of the 690-session archive. \\
\rowcolor{black!5}
Archived forms & 180 sessions & 30 paired live runs and 150 released sessions (11 human, 139 default-agent); baseline form evaluation. \\
\rowcolor{white}
Additional form runs & 10 sessions & Paired no-pause \DarkAgent runs over five seeds and two forms (190 form sessions total). \\
\rowcolor{black!5}
White-box honeysite audit & 172 sessions & 24 stock Chromium, 36 headful nodriver Chrome, 48 \Browsium, and 64 external stealth baseline sessions. \\
\rowcolor{white}
\FPAgent reference & 7,719 records & 546 human and 7,173 agent records with complete attributes; environment-model fitting and evaluation. \\
\rowcolor{black!5}
Stock-browser controls & 14 launches & Windows browser launches with no violations across 29 runtime checks. \\
\bottomrule
\end{tabularx}
\end{table*}

Table~\ref{tab:mimictrace-cohorts} summarizes each cohort and its experimental
role. The 11 released AgentShield human form sessions come from six authors
across four source-address groups, which do not uniquely identify participants.
In the packaged archive, human sessions are labeled \texttt{train} (8) and
\texttt{calibration} (3), while all agent sessions reside in \texttt{test};
these archive labels support frozen-detector packaging and do not imply
participant-disjoint folds in the repeated supervised form C2ST.

\subsection{Behavioral Views and Statistical Scope}
\label{app:mimictrace-behavior}

On the form suite (pointer events throttled above $40$\,ms), the per-channel
C2STs use 10 keyboard features, 9 mouse features, and 5 pacing features. The
pre-click pause policy in Table~\ref{tab:mimictrace-summary} samples from
$3{,}110$ click movements across 43 browser-disjoint \FPAgent human groups
($47.2$\,ms training median) without touching AgentShield human sessions. Two
statistical bounds apply to Table~\ref{tab:mimictrace-summary}: the reported
permutation $p$-values test each configuration separately against Humanize or
against a 200-shuffle label null rather than testing the paired before/after
pause difference directly, and non-significance against the label null
($p=0.264$ on pacing) does not establish distribution equality. The small
sample ($11$ vs.\ $10$) limits interpretation.

Similarly, cross-dataset differences (such as frozen \MouseGen scoring $0.7197$
on $9$\,ms Bo\u{g}azi\c{c}i trajectories versus $0.9916$ on $\approx 16.4$\,ms
\FPAgent browsing traces, or \KeyboardGen scoring $0.6809$ on Aalto sentences
versus $0.646$--$0.680$ on short form fields) change both task and recorder and
therefore do not isolate a single factor. Within the three-user Bo\u{g}azi\c{c}i
test split, one user (\texttt{bogazici\_16}) contributes 32 of the $4{,}000$
test pairs, so the pooled $0.7197$ ROC-AUC primarily reflects the other two
held-out users (per-user range $0.562$--$0.877$).

\subsection{Environment Features, Controls, and Rule Breakdown}
\label{app:mimictrace-environment}

The base environment feature vector comprises 178 features: 42 wire-order HTTP
header features, 95 browser-surface attributes, and 41 cross-channel
consistency features (28 cross-channel checks, 11 container/locale signals, and
two summary counts), with 23 optional human-reference novelty digests; IP, ASN,
and geolocation fields are excluded.

\noindent\textbf{Header allowlist control.}
Request headers enter through a 15-name browser allowlist. Stripping and
poisoning excluded headers (\texttt{Host}, \texttt{Cookie}, \texttt{X-*},
\texttt{ASN}, and \texttt{City}) leaves extracted features bit-identical across
all $7{,}728$ raw \FPAgent records ($7{,}719$ complete: 546 human, $7{,}173$
agent) and all 172 honeysite sessions. Without this filter, excluded header
names alone yield $1.000$ ROC-AUC on \FPAgent because server-added \texttt{ASN}
and \texttt{City} headers appear only on agent records; allowlisting removes
that artifact without eliminating host-level differences.

\noindent\textbf{Runtime probe and baseline rule breakdown.}
The \texttt{runtime\_integrity} probe evaluates 29 runtime checks across four
groups: (i)~a 15-module WebAssembly feature ladder and the CPU NaN sign of
\texttt{f32.sub(inf, inf)} against \texttt{userAgentData.architecture}, typed
arrays, and engine error dialects; (ii)~native builtin structure via
\texttt{toString}, \texttt{Reflect.construct}, WebIDL brand-check
\texttt{TypeError}s, and call stacks; (iii)~value agreement across the top
window, a fresh hidden \texttt{iframe}, and a web worker; and (iv)~Pointer/UI
Events input properties including active pointer pressure and key codes. Zero
runtime checks flag a violation across 14 stock Windows launches (Chrome~157, Edge~157,
Playwright Chromium~149, Firefox~150) or across the 48 \Browsium sessions
(configured with hardware D3D11 rendering and backend-supported WebGL
extensions), whereas controlled spoofing triggers the expected dialect, ladder,
brand-check, and worker-mismatch rules.

On \FPAgent, logistic regression reaches $0.977$ ROC-AUC under agent-family
holdout; under whole-device holdout, the cross-channel consistency block scores
$0.774$ (LogReg) and $0.742$ (XGBoost), led by stacks with internal
contradictions (ChatGPT Agent $0.984$, Manus $0.984$, Skyvern $0.885$). The
cross-channel checks flag $995/995$ ChatGPT Agent and $959/988$ Manus
sessions on \FPAgent, $12/15$ Camoufox honeysite sessions via foreign-OS font
and GPU mismatches (while the separate \texttt{Accept-Language}/\texttt{Intl}
locale signal fires in $15/15$), and $12/12$ Patchright sessions via unreduced
user-agent strings or missing \texttt{window.chrome} and
\texttt{Accept-Language}. Among imported baselines in
Table~\ref{tab:mimictrace-channels}, the combined OR detector also flags all
12 headless nodriver and all 12 undetected-chromedriver sessions.

\subsection{Transfer to Held-Out Stealth Families}
\label{app:mimictrace-transfer}

In the three-seed leave-one-stealth-family-out evaluation against unseen human
device groups (combining count-free behavioral kinematics/rates with
cross-channel and reference features), holding out \Browsium yields $0/12$
\DarkAgent detections ($0.532$ ROC-AUC), whereas holding out CloakBrowser,
Camoufox, and Patchright yields ROC-AUCs of $0.905$, $0.945$, and $0.962$
(detection rates $33.3\%$, $73.3\%$, and $66.7\%$). Because realized human test
FPR varies from $3.2\%$ to $7.9\%$ across held-out splits, ROC-AUC rather than
raw detection percentage provides the threshold-free comparison across families.
\end{document}